\documentclass[12pt]{article}
\usepackage{enumerate}
\usepackage{amsfonts}
\usepackage{amsmath}
\usepackage{amssymb}

\newcounter{defin}  \newcounter{lemma}  \newcounter{theorem}
\newcounter{property} \newcounter{corol}  \newcounter{remark} \newcounter{example}

\newenvironment{lemma}{\par\refstepcounter{lemma}
     \textbf{Lemma \thelemma.} }{\rm\par}
\newenvironment{theorem}{\par\refstepcounter{theorem}
     \textbf{Theorem \thetheorem.}\ }{\rm\par}
\newenvironment{property}{\par\refstepcounter{property}
     \textbf{Proposition \theproperty.}\ }{\rm\par}
\newenvironment{corollary}{\par\refstepcounter{corol}
     \textbf{Corollary \thecorol.} }{\rm\par}
\newenvironment{definition}{\par\refstepcounter{defin}
     \textbf{Definition \thedefin.}\ }{\rm\par}
\newenvironment{remark}{\par\refstepcounter{remark}
     \textbf{Remark \theremark.}}{\rm\par}
\newenvironment{example}{\par\refstepcounter{example}
     \textbf{Example \theexample.}}{\rm\par}

\begin{document}
\title{Continuity of the von Neumann entropy}
\author{M.E.Shirokov\thanks{e-mail:msh@mi.ras.ru}\\\\Steklov Mathematical Institute, Moscow,
Russia}\date{} \maketitle

\tableofcontents

\pagebreak

\section{Introduction}

The set of quantum states -- density operators in a separable
Hilbert space --  plays the central role in analysis of general
infinite dimensional quantum systems. One of the technical problems
in this analysis is related to noncompactness of the set of quantum
states and nonexistence of inner points of this set considered as a
closed convex subset of the separable Banach space of all
trace\nobreakdash-\hspace{0pt}class operators. Another technical
problem consists in discontinuity and unboundedness of basic
characteristics of quantum states such as the von Neumann entropy,
the relative entropy, etc. The above problems can be partially
overcome by using two special properties of the set of quantum
states considered in detail in the first part of \cite{Sh-9}. The
first of them can be considered as a kind of "weak compactness"
since it provides generalization to the set of quantum states of
several results well known for compact convex sets (see section 2)
while the second one called the stability property reveals the
special relation between the topology and the convex structure of
the set of quantum states (see subsection 3.1). These two properties
provide the foundation of analysis of continuity of several
important characteristics of quantum systems and quantum channels
(see \cite{Sh-9} and the reference therein).

In this paper we prove a stronger version of the stability property
of the set of quantum states naturally called \textit{strong
stability}  and consider  its applications concerning the problem of
approximation of concave (convex) functions on the set of quantum
states and providing the new approach to analysis of continuity of
such functions.

The main application of the strong stability property considered in
this paper is the development of a method of proving continuity of
the von Neumann entropy. In infinite dimensions the von Neumann
entropy is a nonnegative concave lower semicontinuous function on
the set of quantum states taking the value $+\infty$ on a dense
subset of this set.\footnote{Moreover, the set of states with finite
entropy is a first category subset of the set of all quantum states
\cite{W}.} Nevertheless the von Neumann entropy has continuous
bounded restrictions to some important subsets of quantum states,
for example, to the set of states of the system of quantum
oscillators with bounded mean energy. Since continuity of the
entropy is a very desirable property in analysis of quantum systems,
various sufficient continuity conditions have been obtained up to
now. The earliest among them seems to be Simon's dominated
convergence theorems presented in \cite{Simon} and widely used in
applications (the generalized forms of these theorems are presented
in corollary \ref{cont-cond++}). Another useful continuity condition
originally appeared in \cite{W} (as far as I know) and can be
formulated as continuity of the entropy on each subset of states
characterized by bounded mean value of a given positive unbounded
operator with discrete spectrum provided that its sequence of
eigenvalues has a sufficient rate of increase (see example
\ref{bounded-energy}). Some special conditions of continuity of the
von Neumann entropy are considered in  \cite{Sh-4}. It turns out
that the strong stability property of the set of quantum states
(more precisely, the approximation technique based on this property)
provides a new method of proving continuity of the von Neumann
entropy on a set of quantum states based on the established relation
between this property and the special \textit{uniform approximation
property} of this set defined via the relative entropy. Well known
results concerning the relative entropy make it possible to show
conservation of the uniform approximation property under different
set-operations, which implies roughly speaking "preservation of
continuity" of the entropy under these set-operations.

The proposed method makes possible to re-derive the known conditions
of continuity of the von Neumann entropy mentioned above (in the
more general forms) and to obtain the several new (as far as I know)
conditions which seems to be useful in analysis of quantum systems.

\section{Preliminaries}

Let $\mathcal{H}$ be a separable Hilbert space,
$\mathfrak{B}(\mathcal{H})$ -- the Banach space of all bounded
operators in $\mathcal{H}$ with the operator norm $\Vert\cdot\Vert$,
$\mathfrak{T}( \mathcal{H})$ -- the Banach space of all
trace\nobreakdash-\hspace{0pt}class operators in $\mathcal{H}$ with
the trace norm $\Vert\cdot\Vert_{1}$, containing the cone
$\mathfrak{T}_{+}(\mathcal{H})$ of all positive
trace\nobreakdash-\hspace{0pt}class operators. The closed convex
subsets
$$
\mathfrak{T}_{1}(\mathcal{H})=\{A\in\mathfrak{T}_{+}(\mathcal{H})\,|\,\mathrm{Tr}A\leq
1\}\;\;\textup{and}\;\;\mathfrak{S}(\mathcal{H})=\{A\in\mathfrak{T}_{+}(\mathcal{H})\,|\,\mathrm{Tr}A=1\}
$$
are complete separable metric spaces with the metric defined by the
trace norm. Operators in $\mathfrak{S}(\mathcal{H})$ are denoted
$\rho,\sigma,\omega,...$ and called density operators or states
since each density operator uniquely defines a normal state on
$\mathfrak{B}(\mathcal{H})$.

In what follows $\mathcal{A}$ is a subset of the cone of positive
trace\nobreakdash-\hspace{0pt}class operators.

We denote by $\mathrm{cl}(\mathcal{A})$, $\mathrm{co}(\mathcal{A})$,
$\sigma\textrm{-}\mathrm{co}(\mathcal{A})$,
$\overline{\mathrm{co}}(\mathcal{A})$ and
$\mathrm{extr}(\mathcal{A})$ the closure, the convex hull, the
$\sigma$\nobreakdash-\hspace{0pt}convex hull, the convex closure and
the set of all extreme points of a set $\mathcal{A}$ correspondingly
\cite{J&T,R}.

In what follows we consider functions on subsets of
$\mathfrak{T}_{+}(\mathcal{H})$ taking values in
$[-\infty,+\infty]$, which are \textit{semibounded} (either lower or
upper bounded) on these subsets.

We denote by $\mathrm{co}f$ and $\overline{\mathrm{co}}f$ the convex
hull and the convex closure of a function $f$ on a convex set
$\mathcal{A}$ \cite{J&T,R}.

The set of all bounded continuous functions on a set $\mathcal{A}$
is denoted $C(\mathcal{A})$.

The set of all Borel probability measures on a closed set
$\mathcal{A}$ endowed with the topology of weak convergence is
denoted $\mathcal{P}(\mathcal{A})$. This set can be considered as a
complete separable metric space \cite{Par}. The \textit{barycenter}
$\textbf{b}(\mu)$ of the measure $\mu$ in $\mathcal{P}(\mathcal{A})$
is the operator in $\overline{\mathrm{co}}(\mathcal{A})$ defined by
the Bochner integral\footnote{This integral always exists if the set
$\mathcal{A}$ is bounded.}
\[
\textbf{b}(\mu)=\int_{\mathcal{A}}A\mu(dA).
\]

For arbitrary subset
$\,\mathcal{B}\subseteq\overline{\mathrm{co}}(\mathcal{A})\,$ let
$\mathcal{P}_{\mathcal{B}}(\mathcal{A})$ be the subset of
$\mathcal{P}(\mathcal{A})$ consisting of all measures with
barycenter in $\mathcal{B}$.

Let $\mathcal{P}^{\mathrm{a}}(\mathcal{A})$ be the subset of
$\mathcal{P}(\mathcal{A})$ consisting of atomic measures and let
$\mathcal{P}^{\mathrm{f}}(\mathcal{A})$ be the subset of
$\mathcal{P}^{\mathrm{a}}(\mathcal{A})$ consisting of measures with
a finite number of atoms. Each measure in
$\mathcal{P}^{\mathrm{a}}(\mathcal{A})$ corresponds to a collection
of operators $\{A_{i}\}\subset\mathcal{A}$ with probability
distribution $\{\pi_{i}\}$ conventionally called an
\textit{ensemble} and denoted $\{\pi_{i},A_{i}\}$. The barycenter of
this measure is the average $\sum_{i}\pi_{i}A_{i}$ of the
corresponding ensemble.

We use the following two strengthened versions of the well known
notion of a concave function.

A semibounded function $f$ on the set $\mathfrak{S}(\mathcal{H})$ is
called $\sigma$\nobreakdash-\hspace{0pt}\textit{concave} at a state
$\rho_{0}\in\mathfrak{S}(\mathcal{H})$ if the discrete Jensen's
inequality
$$
f(\rho_{0})\geq\sum_{i}\pi_{i}f(\rho_{i})
$$
holds for an arbitrary \textit{countable} ensemble
$\{\pi_{i},\rho_{i}\}$ of states in $\mathfrak{S}(\mathcal{H})$ with
the average state $\rho_{0}$.

A semibounded universally measurable\footnote{This means that the
function $f$ is measurable with respect to any measure in
$\mathcal{P}(\mathfrak{S}(\mathcal{H}))$.} function $f$ on the set
$\mathfrak{S}(\mathcal{H})$ is called
$\mu$\nobreakdash-\hspace{0pt}\textit{concave} at a state
$\rho_{0}\in\mathfrak{S}(\mathcal{H})$ if the integral Jensen's
inequality
$$
f(\rho_{0})\geq\int_{\mathfrak{S}(\mathcal{H})}f(\rho)\mu(d\rho)
$$
holds for an arbitrary measure $\mu$ in
$\mathcal{P}(\mathfrak{S}(\mathcal{H}))$ with the barycenter
$\rho_{0}$.

$\sigma$\nobreakdash-\hspace{0pt}\textit{convexity} and
$\mu$\nobreakdash-\hspace{0pt}\textit{convexity} of a function $f$
are naturally defined via the above notions applied to the function
$-f$.

The examples of semibounded functions (in particular, Borel
functions\footnote{If $\dim\mathcal{H}<+\infty$ then arbitrary
convex Borel function on the set $\mathfrak{S}(\mathcal{H})$ with
the range $[-\infty, +\infty]$ is
$\sigma$\nobreakdash-\hspace{0pt}convex and
$\mu$\nobreakdash-\hspace{0pt}convex at any state \cite{Gus}.}) on
the set $\mathfrak{S}(\mathcal{H})$, which are convex but not
$\sigma$\nobreakdash-\hspace{0pt}convex or
$\sigma$\nobreakdash-\hspace{0pt}convex but not
$\mu$\nobreakdash-\hspace{0pt}convex at particular states as well as
sufficient conditions for $\sigma$\nobreakdash-\hspace{0pt}convexity
and $\mu$\nobreakdash-\hspace{0pt}convexity of a convex function at
any state are considered in \cite{Sh-9}.

The identity operator in a Hilbert space $\mathcal{H}$ and the
identity transformation of the space $\mathfrak{T}(\mathcal{H})$ are
denoted $I_{\mathcal{H}}$ and $\mathrm{Id}_{\mathcal{H}}$
correspondingly.

Following \cite{H-Sh-2} an arbitrary positive unbounded operator in
a Hilbert space with discrete spectrum of finite multiplicity is
called $\mathfrak{H}$\nobreakdash-\hspace{0pt}\textit{operator}.

The set $\mathfrak{S}(\mathcal{H})$ is not compact if
$\dim\mathcal{H}=+\infty$, but it has the property consisting in
compactness of the pre-image
$\textbf{b}^{-1}(\mathcal{A})\subset\mathcal{P}(\mathfrak{S}(\mathcal{H}))$
of any compact subset $\mathcal{A}$ of $\mathfrak{S}(\mathcal{H})$
under the  map $\mu\mapsto\textbf{b}(\mu)$ \cite[Proposition
2]{H-Sh-2}, which can be used for proving for the set
$\mathfrak{S}(\mathcal{H})$ and for its subsets several results well
known for compact sets. This property (in the general context of a
metrizable convex subset of a locally convex space) is studied in
detail in \cite{P-Sh}, where it is called
$\mu$-\textit{compactness}. It implies in particular the following
Choquet-type assertion and the lemma below.\vspace{5pt}
\begin{lemma}\label{ca-prop-1}
\textit{Let $\mathcal{A}$ be a closed subset of
$\mathfrak{S}(\mathcal{H})$.} \textit{For an arbitrary state $\rho$
in $\,\overline{\mathrm{co}}(\mathcal{A})$ there exists a measure
$\mu$ in $\,\mathcal{P}(\mathcal{A})$ such that
$\,\textbf{b}(\mu)=\rho$.}
\end{lemma}\vspace{5pt}

\textbf{Proof.} Let $\rho_{0}\in\overline{\mathrm{co}}(\mathcal{A})$
and $\{\rho_{n}\}\subset\mathrm{co}(\mathcal{A})$ be a sequence
converging to the state $\rho_{0}$. For each $n\in\mathbb{N}$ there
exists  a measure  $\mu_{n}\in\mathcal{P}(\mathcal{A})$ with finite
support such that  $\rho_{n}=\textbf{b}(\mu_{n})$. By
$\mu$-compactness of the set $\mathfrak{S}(\mathcal{H})$ the
sequence $\{\mu_{n}\}$ has a partial limit  $\mu_{0}\in
\mathcal{P}(\mathcal{A})$. Continuity of the map
$\mu\mapsto\textbf{b}(\mu)$ implies $\textbf{b}(\mu_{0})=\rho_{0}$.
$\square$\vspace{5pt}

Lemma \ref{ca-prop-1} provides correctness of the definition of the
functions in the following lemma, proved in the
Appendix.\vspace{5pt}

\begin{lemma}\label{ca-l-2}
\textit{Let $f$ be a lower semicontinuous lower bounded function on
a closed subset $\mathcal{A}$ of
$\,\mathfrak{S}(\mathcal{H})$.}\vspace{5pt}

A) \textit{The function
$$
\check{f}_{\mathcal{A}}(\rho)\mapsto\inf_{\mu\in
\mathcal{P}_{\{\rho\}}(\mathcal{A})}\int_{\mathcal{A}}
f(\sigma)\mu(d\sigma)
$$
is convex and lower semicontinuous on the set
 $\,\overline{\mathrm{co}}(\mathcal{A})$. For arbitrary
$\rho$ in $\overline{\mathrm{co}}(\mathcal{A})$ the infimum in the
definition of the value $\check{f}_{\mathcal{A}}(\rho)$ is achieved
at a particular measure in
$\mathcal{P}_{\{\rho\}}(\mathcal{A})$.}\vspace{5pt}

B) \textit{If the map $\mathcal{P}(\mathcal{A})\ni\mu\mapsto
\textbf{b}(\mu)\in \overline{\mathrm{co}}(\mathcal{A})$ is open then
the function
$$
\hat{f}_{\mathcal{A}}(\rho)=\sup_{\mu\in
\mathcal{P}_{\{\rho\}}(\mathcal{A})}\int_{\mathcal{A}}
f(\sigma)\mu(d\sigma)
$$
is concave and lower semicontinuous on the set
$\,\overline{\mathrm{co}}(\mathcal{A})$.}
\end{lemma}\vspace{5pt}

For given natural $k$ we denote by
$\mathfrak{T}_{+}^{k}(\mathcal{H})$ (correspondingly by
$\mathfrak{S}_{k}(\mathcal{H})$) the set of positive
trace\nobreakdash-\hspace{0pt}class operators (correspondingly
states) having rank $\leq k$. The convex set
$\bigcup_{k=1}^{+\infty}\mathfrak{S}_{k}(\mathcal{H})$ of all finite
rank states is denoted $\mathfrak{S}_{\mathrm{f}}(\mathcal{H})$.

A linear positive trace\nobreakdash-\hspace{0pt}nonincreasing map
$\Phi:\mathfrak{T}(\mathcal{H})\rightarrow\mathfrak{T}(\mathcal{H})$
such that the dual map
$\Phi^{*}:\mathfrak{B}(\mathcal{H})\rightarrow\mathfrak{B}(\mathcal{H})$
is completely positive is called a \textit{quantum operation}
\cite{H-SSQT}. The convex set of all quantum operations from
$\mathfrak{T}(\mathcal{H})$ to itself is denoted $\mathfrak{F}_{\leq
1}(\mathcal{H})$. If a quantum operation $\Phi$ is
trace\nobreakdash-\hspace{0pt}preserving then it is called a
\textit{quantum channel}.

Arbitrary quantum operation (correspondingly channel)
$\Phi\in\mathfrak{F}_{\leq 1}(\mathcal{H})$ has the following Kraus
representation
$$
\Phi(\cdot)=\sum_{j=1}^{+\infty}V_{j}(\cdot)V^{*}_{j},
$$
where $\{V_{j}\}_{j=1}^{+\infty}$ is a set of operators in
$\mathfrak{B}(\mathcal{H})$ such that
$\sum_{j=1}^{+\infty}V^{*}_{j}V_{j}\leq I_{\mathcal{H}}$
(correspondingly
$\sum_{j=1}^{+\infty}V^{*}_{j}V_{j}=I_{\mathcal{H}}$).\vspace{5pt}

For given natural $n$ we denote by $\mathfrak{F}^{n}_{\leq
1}(\mathcal{H})$ the subset of  $\mathfrak{F}_{\leq 1}(\mathcal{H})$
consisting of quantum operations having the Kraus representation
with $\leq n$ nonzero summands.\vspace{5pt}

We will use the following result of the purification
theory.\footnote{The assertion of the below lemma can be proved by
noting that the infimum in the definition of the Bures distance (or
the supremum in the definition of the Uhlmann fidelity) between two
quantum states can be taken only over all purifications of one state
with fixed purification of the another state and that convergence of
a sequence of states in the trace norm distance implies its
convergence in the Bures distance \cite{H-IR,N&Ch}.}\vspace{5pt}
\begin{lemma}\label{purification}
\textit{Let $\mathcal{H}$ and $\mathcal{K}$ be Hilbert spaces such
that $\,\dim\mathcal{H}=\dim\mathcal{K}$. For an arbitrary pure
state $\omega_{0}$ in
$\,\mathfrak{S}(\mathcal{H}\otimes\mathcal{K})$ and an arbitrary
sequence $\{\rho_{n}\}$ of states in $\,\mathfrak{S}(\mathcal{H})$
converging to the state
$\rho_{0}=\mathrm{Tr}_{\mathcal{K}}\omega_{0}$ there exists a
sequence $\{\omega_{n}\}$ of pure states in
$\,\mathfrak{S}(\mathcal{H}\otimes\mathcal{K})$ converging to the
state $\omega_{0}$ such that\break
$\rho_{n}=\mathrm{Tr}_{\mathcal{K}}\omega_{n}$ for all $n$.}
\end{lemma}\vspace{5pt}

Let $\mathfrak{P}_{n}$ be the set of all probability distributions
with $n\leq+\infty$ outcomes endowed with the total variation
topology.\vspace{5pt}

\textbf{Note:}  In what follows  continuity of a function $f$ on a
set $\mathcal{A}\subset\mathfrak{T}_{+}(\mathcal{H})$ implies its
finiteness on this set (in contrast to lower (upper)
semicontinuity).

\section{The strong stability property of $\mathfrak{S}(\mathcal{H})$}

\subsection{The definition}

The notion of stability of a convex subset of a linear topological
space appeared at the end of the 1970's as a result of study of the
properties of compact convex sets, which led in particular to
proving equivalence of continuity of the convex
envelope\footnote{the convex hull in our notations.} of arbitrary
continuous function (the CE\nobreakdash-\hspace{0pt}property),
openness of the mixture map and openness of the barycenter map for
given compact convex set (the Vesterstrom-O'Brien theorem
\cite{Brien}). In the subsequent papers (see
\cite{Grzaslewicz,Susanne} and the reference therein) the term
\textit{stability} was used to denote openness of the mixture map
for arbitrary convex subset of a linear topological space (which is
not equivalent in general to the
CE\nobreakdash-\hspace{0pt}property).

The stability property of the set $\mathfrak{S}(\mathcal{H})$ of
quantum states and its corollaries are considered in detail in
\cite{Sh-9}. It consists in the validity of the following
equivalent\footnote{Equivalence of these statements follows from the
$\mu$-compact generalization of the Vesterstrom-O'Brien theorem
\cite[theorem 1]{P-Sh}.} statements:
\begin{itemize}
    \item the map $\mathfrak{S}(\mathcal{H})^{\times 2}\times[0,1]\ni (\rho,\sigma,\lambda)\ \mapsto\ \lambda  \rho + (1-\lambda)
\sigma \in \mathfrak{S}(\mathcal{H})$ is open;
    \item the map
    $\mathcal{P}(\mathfrak{S}(\mathcal{H}))\ni\mu\mapsto\textbf{b}(\mu)\in\mathfrak{S}(\mathcal{H})$ is
    open;
    \item the map
    $\mathcal{P}(\mathrm{extr}\mathfrak{S}(\mathcal{H}))\ni\mu\mapsto\textbf{b}(\mu)\in\mathfrak{S}(\mathcal{H})$ is
    open;
    \item $\mathrm{co}f=\overline{\mathrm{co}}f\in C(\mathfrak{S}(\mathcal{H}))\;$ for arbitrary $f\,\in
    C(\mathfrak{S}(\mathcal{H}))$;
    \item $f_{*}^{\sigma}=f_{*}^{\mu}\in C(\mathfrak{S}(\mathcal{H}))\;$ for arbitrary $\,f\in
    C(\mathrm{extr}\mathfrak{S}(\mathcal{H}))$, where
    $f_{*}^{\sigma}$ and $f_{*}^{\mu}$ are the
    $\sigma$\nobreakdash-\hspace{0pt}convex roof and the
    $\mu$\nobreakdash-\hspace{0pt}convex roof of the function $f$ \cite{Sh-9}.
 \end{itemize}

Physically openness of the map
$\mathcal{P}(\mathfrak{S}(\mathcal{H}))\ni\mu\mapsto\textbf{b}(\mu)\in\mathfrak{S}(\mathcal{H})$
(correspondingly of the map
$\mathcal{P}(\mathrm{extr}\mathfrak{S}(\mathcal{H}))\ni\mu\mapsto\textbf{b}(\mu)\in\mathfrak{S}(\mathcal{H})$)
means roughly speaking that any small perturbation of the average
state of a given continuous ensemble of states (correspondingly of
pure states) can be realized by appropriate small perturbations of
the states of this ensemble.

It turns out that the stability property of the set
$\mathfrak{S}(\mathcal{H})$ can be strengthened by showing that any
small perturbation of the average state of a given (countable or
continuous) ensemble of finite rank states can be realized by
appropriate small perturbations of the states of this ensemble
\textit{without increasing of the maximal rank of these states}.
Mathematically this \textit{strong stability property} of the set
$\mathfrak{S}(\mathcal{H})$ is formulated in the following
theorem.\vspace{5pt}

\begin{theorem}\label{s-stability}
\textit{The surjective maps
$\;\mathcal{P}(\mathfrak{S}_{k}(\mathcal{H}))\ni\mu\mapsto\textbf{b}(\mu)\in\mathfrak{S}(\mathcal{H})$
and
$\;\mathcal{P}^{\mathrm{a}}(\mathfrak{S}_{k}(\mathcal{H}))\ni\mu\mapsto\textbf{b}(\mu)\in\mathfrak{S}(\mathcal{H})$
are open for each natural
$k$.}\footnote{$\mathfrak{S}_{k}(\mathcal{H})=\{\rho\in\mathfrak{S}(\mathcal{H})\,|\,\mathrm{rank}\rho\leq
k\}$, see section 2.}
\end{theorem}
\vspace{5pt}

As mentioned before the assertion of theorem \ref{s-stability} for
$k=1$ is equivalent to openness of the map
$\mathcal{P}(\mathfrak{S}(\mathcal{H}))\ni\mu\mapsto\textbf{b}(\mu)\in\mathfrak{S}(\mathcal{H})$.
The proof of this equivalence is based on coincidence of the set
$\mathfrak{S}_{1}(\mathcal{H})$ with the set
$\mathrm{extr}\mathfrak{S}(\mathcal{H})$ and is universal in the
sense that it is valid for arbitrary compact or
$\mu$\nobreakdash-\hspace{0pt}compact convex set in the role of
$\mathfrak{S}(\mathcal{H})$ \cite{Brien, P-Sh}. In contrast to this
in the proof of the assertion of theorem \ref{s-stability} for $k>1$
the specific structure of the set $\mathfrak{S}(\mathcal{H})$ is
essentially used.\vspace{5pt}

The basic ingredients of the proof of the above theorem are the
following lemma and lemma \ref{density-k} below.\vspace{5pt}

\begin{lemma}\label{twist}
\textit{Let $\{\pi^{0}_{i}, \rho^{0}_{i}\}$ be a countable ensemble
of states in $\,\mathfrak{S}_{k}(\mathcal{H})$ with the average
$\rho_{0}=\sum_{i=1}^{+\infty}\pi^{0}_{i}\rho^{0}_{i}$. For an
arbitrary sequence $\{\rho_{n}\}\subset\mathfrak{S}(\mathcal{H})$
converging to the state $\rho_{0}$ there exists a sequence
$\{\{\pi^{n}_{i}, \rho^{n}_{i}\}\}_{n}$ of countable ensembles of
states in $\,\mathfrak{S}_{k}(\mathcal{H})$ such that}
$$
\lim_{n\rightarrow+\infty}\pi^{n}_{i}=\pi^{0}_{i},\quad
\pi^{0}_{i}>0\;\Rightarrow
\lim_{n\rightarrow+\infty}\rho^{n}_{i}=\rho^{0}_{i},\;\; \forall
i,\quad \textit{and} \quad
\rho_{n}=\sum_{i=1}^{+\infty}\pi^{n}_{i}\rho^{n}_{i},\;\;\forall n.
$$
\end{lemma}\vspace{5pt}

The assertion of this lemma implies weak convergence of the sequence
$\{\{\pi^{n}_{i}, \rho^{n}_{i}\}\}_{n}$ of atomic measures to the
atomic measure $\{\pi^{0}_{i}, \rho^{0}_{i}\}$, t.i. convergence in
$\mathcal{P}(\mathfrak{S}_{k}(\mathcal{H}))$, which means that
$\lim_{n\rightarrow+\infty}\sum_{i}\pi^{n}_{i}f(\rho^{n}_{i})=\sum_{i}\pi^{0}_{i}f(\rho^{0}_{i})$
for any function $f$ in $C(\mathfrak{S}_{k}(\mathcal{H}))$. This
relation can be easily proved by noting that pointwise convergence
of the sequence $\{\{\pi_{i}^{n}\}\}_{n}$ to the probability
distribution $\{\pi_{i}^{0}\}$ implies its convergence in the norm
of total variation.\vspace{5pt}

\textbf{Proof of lemma \ref{twist}.} For each $i$ let
$|\varphi_{i}\rangle$ be a unit vector in
$\mathfrak{S}(\mathcal{H}\otimes\mathcal{H}_{k})$ such that
$\mathrm{Tr}_{\mathcal{H}_{k}}|\varphi_{i}\rangle\langle\varphi_{i}|=\rho^{0}_{i}$,
where $\mathcal{H}_{k}$ is an auxiliary $k$-dimensional Hilbert
space. Let $\{|\epsilon_{i}\rangle\}_{i=1}^{+\infty}$ be an
orthonormal basis in a separable Hilbert space $\mathcal{H}'$.
Consider the unit vector
$|\psi_{0}\rangle=\sum_{i=1}^{+\infty}\sqrt{\pi_{i}^{0}}|\varphi_{i}\rangle\otimes|\epsilon_{i}\rangle$
in the space $\mathcal{H}\otimes\mathcal{H}_{k}\otimes\mathcal{H}'$.
It is easy to see that
$\mathrm{Tr}_{\mathcal{H}_{k}\otimes\mathcal{H}'}|\psi_{0}\rangle\langle\psi_{0}|=\rho_{0}$.
By lemma \ref{purification} there exists sequence
$\{|\psi_{n}\rangle\}$ of unit vectors in
$\mathcal{H}\otimes\mathcal{H}_{k}\otimes\mathcal{H}'$ converging to
the vector $|\psi_{0}\rangle$ such that
$\mathrm{Tr}_{\mathcal{H}_{k}\otimes\mathcal{H}'}|\psi_{n}\rangle\langle\psi_{n}|=\rho_{n}$
for each $n$.

Let $\{E_{i}=I_{\mathcal{H}}\otimes I_{\mathcal{H}_{k}}\otimes
|\epsilon_{i}\rangle\langle\epsilon_{i}|\}_{i=1}^{+\infty}$ be the
local measurement in the space
$\mathcal{H}\otimes\mathcal{H}_{k}\otimes\mathcal{H}'$
\cite{H-SSQT}. Since
$E_{i}|\psi_{0}\rangle=\sqrt{\pi_{i}^{0}}|\varphi_{i}\rangle\otimes|\epsilon_{i}\rangle$
for each $i$ we have \break
$\pi_{i}^{0}=\mathrm{Tr}E_{i}|\psi_{0}\rangle\langle \psi_{0}|$ and
$\pi_{i}^{0}\rho^{0}_{i}=\mathrm{Tr}_{\mathcal{H}_{k}\otimes\mathcal{H}'}E_{i}|\psi_{0}\rangle\langle
\psi_{0}|E_{i}$. Let
$\pi_{i}^{n}=\mathrm{Tr}E_{i}|\psi_{n}\rangle\langle \psi_{n}|$ and
$$
\rho^{n}_{i}=\left\{
   \begin{array}{ll}
    (\pi_{i}^{n})^{-1}\mathrm{Tr}_{\mathcal{H}_{k}\otimes\mathcal{H}'}E_{i}|\psi_{n}\rangle\langle
\psi_{n}|E_{i}, & \pi_{i}^{n}>0\\
    \rho^{0}_{i}, & \pi_{i}^{n}=0.
    \end{array}\right.
$$
Then $\mathrm{rank}\rho^{n}_{i}\leq k$ for all $n$ and $i$. The
sequence of ensembles $\{\pi^{n}_{i}, \rho^{n}_{i}\}$ has the
required properties. $\square$\vspace{5pt}

\begin{remark}\label{twist+}
It is interesting to compare the above lemma with the lemma 3 in
\cite{Sh-2} containing the analogous assertion concerning finite
ensembles with no rank restriction on states of ensembles. The case
of finite ensemble $\{\pi^{0}_{i}, \rho^{0}_{i}\}_{i=1}^{m}$ is
naturally embedded in the condition of lemma \ref{twist} by setting
$\pi^{0}_{i}=0$ for all $i>m$ but this lemma does not guarantee that
the sequence $\{\{\pi^{n}_{i}, \rho^{n}_{i}\}\}_{n}$  consists of
ensembles of $m$ states in contrast to the assertion of lemma 3 in
\cite{Sh-2}. Increasing dimensionality of ensembles of the sequence
$\{\{\pi^{n}_{i}, \rho^{n}_{i}\}\}_{n}$ is the cost of the rank
restriction on the states of these ensembles.$\square$
\end{remark}\vspace{5pt}

For arbitrary state $\rho$ in $\mathfrak{S}(\mathcal{H})$ the set
$\mathcal{P}_{\{\rho\}}^{\mathrm{a}}(\mathfrak{S}(\mathcal{H}))$ is
a dense subset of
$\mathcal{P}_{\{\rho\}}(\mathfrak{S}(\mathcal{H}))$ \cite[lemma
1]{H-Sh-2}. This simple result can be strengthened as
follows.\vspace{5pt}

\begin{lemma}\label{density-k}
\textit{For arbitrary state $\rho$ in $\,\mathfrak{S}(\mathcal{H})$
and $k\in\mathbb{N}$ the set
$\mathcal{P}_{\{\rho\}}^{\mathrm{a}}(\mathfrak{S}_{k}(\mathcal{H}))$
is a dense subset of
$\mathcal{P}_{\{\rho\}}(\mathfrak{S}_{k}(\mathcal{H}))$. }
\end{lemma}\vspace{5pt}

This means that any probability measure supported by the set of
states of rank $\leq k$ can be weakly approximated by some sequence
of atomic measures -- countable ensembles of states of rank $\leq k$
with the same barycenter.

\textbf{Proof.} To prove the assertion of the lemma for $k=1$
consider the Choquet ordering on the set
$\mathcal{P}(\mathfrak{S}(\mathcal{H}))$. We say that $\mu\succ\nu$
if and only if
$$
\int_{\mathfrak{S}(\mathcal{H})}f(\sigma)\mu(d\sigma)
\geq\int_{\mathfrak{S}(\mathcal{H})}f(\sigma)\nu(d\sigma)
$$
for arbitrary convex continuous bounded function $f$ on the set
$\mathfrak{S}(\mathcal{H})$ \cite{Edgar}.

By lemma 1 in \cite{H-Sh-2} for given measure $\mu_{0}$ in
$\mathcal{P}(\mathfrak{S}_{1}(\mathcal{H}))$ there exists a sequence
$\{\mu_{n}\}$ of measures in
$\mathcal{P}(\mathfrak{S}(\mathcal{H}))$ with finite support
converging to the measure $\mu_{0}$ such that
$\mathbf{b}(\mu_{n})=\mathbf{b}(\mu_{0})$ for all $n$. Decomposing
each atom of the measure $\mu_{n}$ into convex combination of pure
states we obtain the measure $\hat{\mu}_{n}$ in
$\mathcal{P}^{\mathrm{a}}(\mathfrak{S}_{1}(\mathcal{H}))$ with the
same barycenter. It is easy to see that $\hat{\mu}_{n}\succ\mu_{n}$.
By $\mu$-compactness of the set $\mathfrak{S}(\mathcal{H})$ the set
$\{\hat{\mu}_{n}\}_{n>0}$ is a relatively compact. This implies
existence of subsequence $\{\hat{\mu}_{n_{k}}\}$ converging to a
measure $\{\hat{\mu}_{0}\}$ in
$\mathcal{P}(\mathfrak{S}_{1}(\mathcal{H}))$ \cite[theorem
6.1]{Par}. Since $\hat{\mu}_{n_{k}}\succ\mu_{n_{k}}$ for all $k$,
the definition of the weak convergence implies
$\hat{\mu}_{0}\succ\mu_{0}$ and hence $\hat{\mu}_{0}=\mu_{0}$ by
maximality of the measure $\mu_{0}$ with respect to the Choquet
ordering \cite{Max}. Density of the set
$\mathcal{P}_{\{\rho\}}^{\mathrm{a}}(\mathfrak{S}_{1}(\mathcal{H}))$
in $\mathcal{P}_{\{\rho\}}(\mathfrak{S}_{1}(\mathcal{H}))$ is
proved. \vspace{5pt}

Let $k>1$ and $\mathcal{H}_{k}$ be the $k$-dimensional Hilbert
space. Let $\Pi$ be the multi-valued map from
$\mathfrak{S}_{k}(\mathcal{H})$ into the set
$2^{\mathfrak{S}_{1}(\mathcal{H}\otimes\mathcal{H}_{k})}$ such that
$\Pi(\rho)$ is the set of all purifications in
$\mathfrak{S}_{1}(\mathcal{H}\otimes\mathcal{H}_{k})$ of the state
$\rho\in\mathfrak{S}_{k}(\mathcal{H})$. It is clear that the map
$\Pi$ is closed-valued. Thus by theorem 3.1 in \cite{Wagner} to
prove existence of a measurable selection of the map $\Pi$ it is
sufficient to show weak measurability of this map in terms of
\cite{Wagner}. Let $U$ be an open subset of
$\mathfrak{S}_{1}(\mathcal{H}\otimes\mathcal{H}_{k})$. Then
$\Pi^{-}(U)=\{\rho\in\mathfrak{S}_{k}(\mathcal{H})\,|\,\Pi(\rho)\cap
U\neq\emptyset\}=\Theta(U)$, where
$\Theta(\cdot)=\mathrm{Tr}_{\mathcal{H}_{k}}(\cdot)$ is the affine
(single valued) map from
$\mathfrak{S}(\mathcal{H}\otimes\mathcal{H}_{k})$ onto
$\mathfrak{S}_{k}(\mathcal{H})$. Since the restriction of the map
$\Theta$ to the set
$\mathfrak{S}_{1}(\mathcal{H}\otimes\mathcal{H}_{k})$ is
open,\footnote{This means that for arbitrary state
$\omega_{0}\in\mathfrak{S}_{1}(\mathcal{H}\otimes\mathcal{H}_{k})$
and  sequence $\{\rho_{n}\}\subset\mathfrak{S}_{k}(\mathcal{H})$
converging to the state $\rho_{0}=\Theta(\omega_{0})$ there exist a
subsequence $\{\rho_{n_{k}}\}$ and a sequence
$\{\omega_{k}\}\subset\mathfrak{S}_{1}(\mathcal{H}\otimes\mathcal{H}_{k})$
converging to the state $\omega_{0}$ such that
$\Theta(\omega_{k})=\rho_{n_{k}}$ for all $k$. The last property can
be verified by using the standard arguments of the purification
theory.} the set $\Pi^{-}(U)=\Theta(U)$ is open and hence Borel. As
mentioned before this implies existence of a measurable selection
$\Pi_{*}$ of the map $\Pi$.\footnote{We use this tedious
argumentation since $\dim\mathcal{H}_{k}<\dim\mathcal{H}$ and hence
we can not refer to the general results of the purification theory.}

Let $\nu_{0}=\mu_{0}\circ\Pi_{*}^{-1}$ be the image of the measure
$\mu_{0}$ under the map $\Pi_{*}$. It is clear that
$\nu_{0}\in\mathcal{P}(\mathfrak{S}_{1}(\mathcal{H}\otimes\mathcal{H}_{k}))$.
By the assertion of the lemma for $k=1$ there exists sequence
$\{\nu_{n}\}$ of measures in
$\mathcal{P}^{\mathrm{a}}(\mathfrak{S}_{1}(\mathcal{H}\otimes\mathcal{H}_{k}))$
converging to the measure $\nu_{0}$ such that
$\textbf{b}(\nu_{n})=\textbf{b}(\nu_{0})$ for all $n$. Since
$\Theta\circ\Pi_{*}=\mathrm{Id}_{\mathcal{H}}$ the image
$\nu_{0}\circ\Theta^{-1}$ of the measure $\nu_{0}$ under the map
$\Theta$ coincides with $\mu_{0}$. This and continuity of the map
$\Theta$ imply convergence of the sequence
$\{\mu_{n}=\nu_{n}\circ\Theta^{-1}\}$ of measures in
$\mathcal{P}^{\mathrm{a}}(\mathfrak{S}_{k}(\mathcal{H}))$ to the
measure $\mu_{0}$.  Since the map $\Theta$ is affine we have
$$
\textbf{b}(\mu_{n})=\Theta(\textbf{b}(\nu_{n}))=\Theta(\textbf{b}(\nu_{0}))=\textbf{b}(\mu_{0})
$$
for all $n$. Thus the sequence $\{\mu_{n}\}$ has the required
properties. $\square$\vspace{5pt}

\textbf{Proof of theorem \ref{s-stability}.} By lemma
\ref{density-k} it is sufficient to prove openness of the surjective
map
$\mathcal{P}^{\mathrm{a}}(\mathfrak{S}_{k}(\mathcal{H}))\ni\mu\mapsto\textbf{b}(\mu)\in\mathfrak{S}(\mathcal{H})$
for each natural $k$.

Let $U$ be an arbitrary open subset of
$\mathcal{P}^{\mathrm{a}}(\mathfrak{S}_{k}(\mathcal{H}))$. Suppose
$\textbf{b}(U)$ is not open in $\mathfrak{S}(\mathcal{H})$. Then
there exist a state $\rho_{0}\in\textbf{b}(U)$ and a sequence
$\{\rho_{n}\}$ of states in
$\mathfrak{S}(\mathcal{H})\setminus\textbf{b}(U)$ converging to the
state $\rho_{0}$.

Let $\mu_{0}=\{\pi_{i}^{0}, \rho_{i}^{0}\}$ be a measure in $U$ such
that $\textbf{b}(\mu_{0})=\rho_{0}$. By lemma \ref{twist} (and the
remark after it) there exists a sequence of measures
$\mu_{n}=\{\pi_{i}^{n}, \rho_{i}^{n}\}$ in
$\mathcal{P}^{\mathrm{a}}(\mathfrak{S}_{k}(\mathcal{H}))$ converging
to the measure $\mu_{0}=\{\pi_{i}^{0}, \rho_{i}^{0}\}$ such that
$\textbf{b}(\mu_{n})=\rho_{n}$ for all $n$. Openness of the set $U$
implies $\mu_{n}\in U$ for all sufficiently large $n$ contradicting
to the choice of the sequence $\{\rho_{n}\}$.$\square$

\subsection{Some implications}

In the case  $\dim\mathcal{H}<+\infty$ the convex (concave) roof
extension to the set $\mathfrak{S}(\mathcal{H})$ of a function $f$
on the set of pure states
$\mathfrak{S}_{1}(\mathcal{H})=\mathrm{extr}\mathfrak{S}(\mathcal{H})$
is defined at a mixed state $\rho$ as the minimal (maximal) value of
$\sum_{i}\pi_{i}f(\rho_{i})$ over all decompositions
$\rho=\sum_{i}\pi_{i}\rho_{i}$ of this state into finite convex
combination of pure states \cite{U}. This extension is widely used
in quantum information theory, in particular, in construction of
entanglement monotones \cite{P&V}. The convex (concave) roof
extension has the two natural generalizations to the case
$\dim\mathcal{H}=+\infty$ called in \cite{Sh-9} the
$\sigma$\nobreakdash-\hspace{0pt}convex (concave) roof and the
$\mu$\nobreakdash-\hspace{0pt}convex (concave) roof correspondingly
(the first extension is defined via all decompositions of a state
into countable convex combination of pure states while the second
one -- via all "continuous" decompositions corresponding to Borel
probability measures on the set of pure states with given
barycenter).

Generalizing  the $\sigma$\nobreakdash-\hspace{0pt}concave roof
construction, for given natural $k$ and semibounded function $f$ on
the set $\mathfrak{S}_{k}(\mathcal{H})$ consider the function
$$
\mathfrak{S}(\mathcal{H})\ni
\rho\mapsto\hat{f}_{k}^{\sigma}(\rho)=\sup_{\{\pi_{i},\rho_{i}\}\in\mathcal{P}^{\mathrm{a}}_{\{\rho\}}(\mathfrak{S}_{k}(\mathcal{H}))}
\sum_{i}\pi_{i}f(\rho_{i})
$$
(the supremum is over all decompositions of the state $\rho$ into
countable convex combination of states of rank $\leq k$). This
function is obviously $\sigma$\nobreakdash-\hspace{0pt}concave on
the set $\mathfrak{S}(\mathcal{H})$ (see section 2). If the function
$f$ is $\sigma$\nobreakdash-\hspace{0pt}concave at any state in
$\mathfrak{S}_{k}(\mathcal{H})$ then the functions
$\hat{f}_{k}^{\sigma}$ and $f$ coincide on the set
$\mathfrak{S}_{k}(\mathcal{H})$, so in this case the function
$\hat{f}_{k}^{\sigma}$ can be considered as an extension of the
function $f$ to the set $\mathfrak{S}(\mathcal{H})$.

Generalizing the $\mu$\nobreakdash-\hspace{0pt}concave roof
construction, for given natural $k$ and semibounded Borel function
$f$ on the set $\mathfrak{S}_{k}(\mathcal{H})$ consider the function
$$
\mathfrak{S}(\mathcal{H})\ni
\rho\mapsto\hat{f}_{k}^{\mu}(\rho)=\sup_{\mu\in\mathcal{P}_{\{\rho\}}(\mathfrak{S}_{k}(\mathcal{H}))}
\int_{\mathfrak{S}_{k}(\mathcal{H})}f(\sigma)\mu(d\sigma)
$$
(the supremum is over all probability measures with the barycenter
$\rho$ supported by states of rank $\leq k$). This function is also
obviously $\sigma$\nobreakdash-\hspace{0pt}concave on the set
$\mathfrak{S}(\mathcal{H})$ but its $\mu$-concavity depends on the
question of its universal measurability.\footnote{By using the
results in \cite{Ressel} it can be proved for bounded function $f$.}
By propositions \ref{k-approx-ls} and \ref{k-approx-us} below the
function $\hat{f}_{k}^{\mu}$ is
$\mu$\nobreakdash-\hspace{0pt}concave on the set
$\mathfrak{S}(\mathcal{H})$ if the function $f$ is either lower
bounded lower semicontinuous or upper bounded upper semicontinuous
on the set $\mathfrak{S}_{k}(\mathcal{H})$. If the function $f$ is
$\mu$\nobreakdash-\hspace{0pt}concave at any state in
$\mathfrak{S}_{k}(\mathcal{H})$ then the functions
$\hat{f}_{k}^{\mu}$ and $f$ coincide on the set
$\mathfrak{S}_{k}(\mathcal{H})$, so in this case the function
$\hat{f}_{k}^{\mu}$ can be considered as an extension of the
function $f$ to the set $\mathfrak{S}(\mathcal{H})$.

The strong stability property of the set $\mathfrak{S}(\mathcal{H})$
stated in theorem \ref{s-stability} and lemma \ref{density-k} imply
the following result.\vspace{5pt}\vspace{5pt}

\begin{property}\label{k-approx-ls}
\textit{Let $f$ be a lower semicontinuous lower bounded function on
the set $\,\mathfrak{S}_{k}(\mathcal{H})$. Then
$\hat{f}_{k}^{\sigma}=\hat{f}_{k}^{\mu}$ and this function is lower
semicontinuous and $\mu$\nobreakdash-\hspace{0pt}concave on the set
$\,\mathfrak{S}(\mathcal{H})$.}
\end{property}\vspace{5pt}

\textbf{Proof.} Coincidence of the functions $\hat{f}_{k}^{\sigma}$
and $\hat{f}_{k}^{\mu}$ follows from lower semicontinuity of the
functional
$\mathcal{P}(\mathfrak{S}_{k}(\mathcal{H}))\ni\mu\mapsto\int_{\mathfrak{S}_{k}(\mathcal{H})}f(\sigma)\mu(d\sigma)$
(proved by the standard argumentation) and lemma \ref{density-k}.
Theorem \ref{s-stability} and lemma \ref{ca-l-2} imply lower
semicontinuity of the lower bounded function
$\hat{f}_{k}^{\sigma}=\hat{f}_{k}^{\mu}$, which guarantees its
$\mu$-concavity (by proposition A-2 in the Appendix in \cite{Sh-9}).
$\square$\vspace{5pt}

The $\mu$-compactness property of the set
$\mathfrak{S}(\mathcal{H})$ (described before lemma \ref{ca-prop-1})
implies the following result.\pagebreak

\begin{property}\label{k-approx-us}
\textit{Let $f$ be an upper semicontinuous upper bounded function on
the set $\,\mathfrak{S}_{k}(\mathcal{H})$. Then the function
$\hat{f}_{k}^{\mu}$ is upper semicontinuous and
$\mu$\nobreakdash-\hspace{0pt}concave on the set
$\,\mathfrak{S}(\mathcal{H})$.}

\textit{For arbitrary state $\rho$ in $\mathfrak{S}(\mathcal{H})$
the supremum in the definition of the value
$\hat{f}_{k}^{\mu}(\rho)$ is achieved at some measure in
$\mathcal{P}_{\{\rho\}}(\mathfrak{S}_{k}(\mathcal{H}))$. }
\end{property}\vspace{5pt}

\textbf{Proof.} Lemma \ref{ca-l-2} implies attainability of the
supremum in the definition of the value $\hat{f}_{k}^{\mu}(\rho)$
and upper semicontinuity of the function $\hat{f}_{k}^{\mu}$, which
guarantees its $\mu$\nobreakdash-\hspace{0pt}concavity (by
proposition A-2 in the Appendix in \cite{Sh-9}).
$\square$\vspace{5pt}

Under the condition of proposition \ref{k-approx-us} we can say
nothing about upper semicontinuity and $\mu$-concavity of the
function $\hat{f}_{k}^{\sigma}$ (see example 2 in
\cite{Sh-9}).\vspace{5pt}

The above two propositions have the obvious corollary.\vspace{5pt}

\begin{corollary}\label{k-approx-c}
\textit{Let $f$ be a  continuous bounded function on the set
$\,\mathfrak{S}_{k}(\mathcal{H})$. Then
$\hat{f}_{k}^{\sigma}=\hat{f}_{k}^{\mu}$ and this function is
continuous on the set $\,\mathfrak{S}(\mathcal{H})$.}
\end{corollary}

\section{On approximation of concave (convex)\\ functions on $\mathfrak{S}(\mathcal{H})$}

The functional constructions considered in subsection 3.2 can be
used in study of the following \textit{approximation problem}: for
given concave (convex) function $f$ on the set
$\mathfrak{S}(\mathcal{H})$ having some particular
symmetry\footnote{This means that the function $f$ is invariant with
respect to the particular family of symmetries of the set
$\mathfrak{S}(\mathcal{H})$.} to find a monotonic sequence
$\{f_{k}\}$ of concave (convex) functions on the set
$\mathfrak{S}(\mathcal{H})$ having the same symmetry, satisfying
additional analytical requirements and such that
$$
f_{k}|_{\mathfrak{S}_{k}(\mathcal{H})}=f|_{\mathfrak{S}_{k}(\mathcal{H})},\;\forall
k, \quad \textrm{and} \quad
\lim_{k\rightarrow+\infty}f_{k}(\rho)=f(\rho),\;\forall
\rho\in\mathfrak{S}(\mathcal{H}).
$$

Let $f$ be a function on the set $\mathfrak{S}(\mathcal{H})$ having
semibounded restriction to the set $\mathfrak{S}_{k}(\mathcal{H})$
for each $k$. We can consider the nondecreasing sequence
$\{\hat{f}_{k}^{\sigma}\}$ of concave functions on the set
$\mathfrak{S}(\mathcal{H})$ and its pointwise limit
$\hat{f}_{+\infty}^{\sigma}=\sup_{k}\hat{f}_{k}^{\sigma}$. If the
restriction of the function $f$ to the set
$\mathfrak{S}_{k}(\mathcal{H})$ is  universally measurable for each
$k$ then we can also consider the nondecreasing sequence
$\{\hat{f}_{k}^{\mu}\}$ of concave functions on the set
$\mathfrak{S}(\mathcal{H})$ and its pointwise limit
$\hat{f}_{+\infty}^{\mu}=\sup_{k}\hat{f}_{k}^{\mu}$.

By construction all the functions in the sequences
$\{\hat{f}_{k}^{\sigma}\}$ and $\{\hat{f}_{k}^{\mu}\}$ inherit the
arbitrary symmetry of the function $f$. Hence the same assertion
holds for the functions $\hat{f}_{+\infty}^{\sigma}$ and
$\hat{f}_{+\infty}^{\mu}$.

The functions $\hat{f}_{+\infty}^{\sigma}$ and
$\hat{f}_{+\infty}^{\mu}$ are concave on the set
$\mathfrak{S}(\mathcal{H})$. By construction
$\hat{f}_{+\infty}^{\sigma}\leq\hat{f}_{+\infty}^{\mu}$ and
$f|_{\mathfrak{S}_{\mathrm{f}}(\mathcal{H})}\leq
\hat{f}_{+\infty}^{\sigma}|_{\mathfrak{S}_{\mathrm{f}}(\mathcal{H})}$
($\mathfrak{S}_{\mathrm{f}}(\mathcal{H})$ is the convex subset of
$\mathfrak{S}(\mathcal{H})$ consisting of finite rank states). If
the function $f$ is $\sigma$\nobreakdash-\hspace{0pt}concave on the
set $\mathfrak{S}(\mathcal{H})$ then $\hat{f}_{+\infty}^{\sigma}\leq
f$, if the function $f$ is $\mu$\nobreakdash-\hspace{0pt}concave on
the set $\mathfrak{S}(\mathcal{H})$ then
$\hat{f}_{+\infty}^{\mu}\leq f$. To show coincidence of the
functions $\hat{f}_{+\infty}^{\sigma}$ and $\hat{f}_{+\infty}^{\mu}$
with the function $f$  additional conditions are
required.\vspace{5pt}

\begin{property}\label{a-problem}
\textit{Let $f$ be a concave lower semicontinuous lower bounded
function on the set $\,\mathfrak{S}(\mathcal{H})$, having some
particular symmetry.}\vspace{5pt}

A) \textit{For each natural $k$ the concave lower semicontinuous
function $\hat{f}_{k}^{\sigma}=\hat{f}_{k}^{\mu}$  has the same
symmetry and coincides with the function $f$ on the set
$\,\mathfrak{S}_{k}(\mathcal{H})$.\break The pointwise limit
$\hat{f}_{+\infty}^{\sigma}=\hat{f}_{+\infty}^{\mu}$ of the
monotonic sequence $\{\hat{f}_{k}^{\sigma}=\hat{f}_{k}^{\mu}\}$
coincides with the function $f$ on the set
$\,\mathfrak{S}(\mathcal{H})$.}\vspace{5pt}

B) \textit{If the function  $f$ has continuous restriction to the
set $\,\mathfrak{S}_{k}(\mathcal{H})$ for each natural $k$ then the
sequence $\,\{\hat{f}_{k}^{\sigma}=\hat{f}_{k}^{\mu}\}$ consists of
concave continuous bounded functions on the set
$\,\mathfrak{S}(\mathcal{H})$.}
\end{property}\vspace{5pt}
\textbf{Proof.} By proposition \ref{k-approx-ls}
$\hat{f}_{k}^{\sigma}=\hat{f}_{k}^{\mu}$ and this function is lower
semicontinuous for each $k$. This implies
$\hat{f}_{+\infty}^{\sigma}=\hat{f}_{+\infty}^{\mu}$ and lower
semicontinuity of the last function. Since the function $f$ is
$\mu$\nobreakdash-\hspace{0pt}concave by proposition A-2 in the
Appendix in \cite{Sh-9}, the first assertion of the proposition
follows from the previous observations and lemma \ref{a-lemma}
below.

The second assertion of the proposition follows from corollary
\ref{k-approx-c} since it is easy to see that continuity of the
restrictions of the concave function $f$ to the set
$\,\mathfrak{S}_{k}(\mathcal{H})$ for all $k$ implies boundedness of
these restrictions. $\square$\vspace{5pt}

\begin{lemma}\label{a-lemma}
\textit{A lower semicontinuous lower bounded concave function $f$ on
the set $\,\mathfrak{S}(\mathcal{H})$ is uniquely determined by its
restriction to the set $\,\mathfrak{S}_{\mathrm{f}}(\mathcal{H})$ of
finite rank states.}
\end{lemma}\vspace{5pt}
\textbf{Proof.} It is sufficient to consider the case of a
nonnegative function $f$.

Let $\rho_{0}$ be an arbitrary state and let
$\left\{\rho_{n}=(\mathrm{Tr}P_{n}\rho_{0})^{-1}P_{n}\rho_{0}\right\}$
be the sequence of finite rank states converging to the state
$\rho_{0}$, where $\{P_{n}\}$ is the sequence of finite rank
spectral projectors of the state $\rho_{0}$ increasing to the
identity operator $I_{\mathcal{H}}$.

For each $n$ the inequality $\lambda_{n}\rho_{n}\leq\rho_{0}$ with
$\lambda_{n}=\mathrm{Tr}P_{n}\rho_{0}$  implies decomposition
$\rho_{0}=\lambda_{n}\rho_{n}+(1-\lambda_{n})\sigma_{n}$, where
$\sigma_{n}=(1-\lambda_{n})^{-1}(\rho-\lambda_{n}\rho_{n})$ is a
state. By concavity and nonnegativity of the function $f$ we have
$f(\rho_{0})\geq\lambda_{n}f(\rho_{n})$ for all $n$, which implies
$\limsup_{n\rightarrow+\infty}f(\rho_{n})\leq f(\rho_{0})$. By lower
semicontinuity of the function $f$ we have
$\lim_{n\rightarrow+\infty}f(\rho_{n})=f(\rho_{0})$. $\square$
\vspace{5pt}

\begin{remark}\label{e-problem}
The first assertion of proposition \ref{a-problem} can be considered
as a "constructive form" of lemma \ref{a-lemma} since it provides a
constructive way of restoring a lower semicontinuous lower bounded
concave function on the set $\mathfrak{S}(\mathcal{H})$ by means of
its restriction to the set $\mathfrak{S}_{\mathrm{f}}(\mathcal{H})$.

Note that the above functions $\hat{f}_{+\infty}^{\sigma}$ and
$\hat{f}_{+\infty}^{\mu}$ can be used in study of the following
\textit{construction problem}: for a given concave function defined
on the convex set $\mathfrak{S}_{\mathrm{f}}(\mathcal{H})$ of finite
rank states and having some particular analytical and  symmetry
properties to construct its concave extension to the set
$\mathfrak{S}(\mathcal{H})$ of all states preserving these
properties. Since in the proof of proposition \ref{a-problem} the
restriction of the function $f$ to the set
$\mathfrak{S}_{\mathrm{f}}(\mathcal{H})$ is only used, it shows that
\textit{for arbitrary concave lower bounded function $f$ on the set
$\,\mathfrak{S}_{\mathrm{f}}(\mathcal{H})$ with some particular
symmetry such that its restriction to the set
$\,\mathfrak{S}_{k}(\mathcal{H})$ is lower semicontinuous for each
$k$ there exists a unique concave lower semicontinuous extension
$\hat{f}_{+\infty}^{\sigma}=\hat{f}_{+\infty}^{\mu}$ to the set
$\,\mathfrak{S}(\mathcal{H})$ with the same symmetry.} For example,
if $f$ is an entropy-type (t.i. nonnegative concave lower
semicontinuous unitary invariant) function defined on the set of
finite rank states then
$\hat{f}_{+\infty}^{\sigma}=\hat{f}_{+\infty}^{\mu}$ is its unique
entropy-type extension to the set of all states. $\square$
\end{remark} \vspace{5pt}

The second assertion of proposition \ref{a-problem} and the
generalized Dini's lemma\footnote{The condition of continuity of the
functions of the increasing sequence in the standard Dini's lemma
can be replaced by the condition of their lower semicontinuity
(provided that the condition of continuity of the limit function is
valid).} imply the following continuity condition.\vspace{5pt}

\begin{corollary}\label{a-problem+} \textit{Let $f$ be a concave lower semicontinuous lower bounded
function on the set $\,\mathfrak{S}(\mathcal{H})$.}\vspace{5pt}

A) \textit{If the function $f$ has continuous restriction to the set
$\,\mathfrak{S}_{k}(\mathcal{H})$ for each $k$ then uniform
convergence of the sequence
$\{\hat{f}_{k}^{\sigma}=\hat{f}_{k}^{\mu}\}$ on a particular subset
$\mathcal{A}\subseteq\mathfrak{S}(\mathcal{H})$ implies continuity
of the function $f$ on this subset.}\vspace{5pt}

B) \textit{Continuity of the function $f$ on a compact subset
$\mathcal{A}\subseteq\mathfrak{S}(\mathcal{H})$ implies uniform
convergence of the sequence
$\{\hat{f}_{k}^{\sigma}=\hat{f}_{k}^{\mu}\}$ on this subset.}

\end{corollary}\vspace{5pt}

We will use corollary \ref{a-problem+} in the next section to obtain
continuity conditions for the von Neumann entropy.

\section{The approximation of the von Neumann\\ entropy and the continuity conditions}

The von Neumann entropy $H(\rho)=-\mathrm{Tr}\rho\log\rho$ is a
lower semicontinuous concave unitary invariant function on the set
$\mathfrak{S}(\mathcal{H})$ of quantum states with the range
$[0,+\infty]$, having continuous restriction to the set
$\mathfrak{S}_{k}(\mathcal{H})$ for each $k$. By proposition
\ref{a-problem} the function $H$ is a pointwise limit of the
increasing sequence $\{H_{k}\}$ of nonnegative concave continuous
bounded\footnote{It is easy to see that the range of the function
$H_{k}$ coincides with $[0,\log k]$.} unitary invariant functions on
the set $\mathfrak{S}(\mathcal{H})$ defined as follows
$$
H_{k}(\rho)=\sup_{\{\pi_{i},\rho_{i}\}\in\mathcal{P}^{\mathrm{a}}_{\{\rho\}}(\mathfrak{S}_{k}(\mathcal{H}))}
\sum_{i}\pi_{i}H(\rho_{i})=\sup_{\mu\in\mathcal{P}_{\{\rho\}}(\mathfrak{S}_{k}(\mathcal{H}))}
\int_{\mathfrak{S}_{k}(\mathcal{H})}H(\sigma)\mu(d\sigma),
$$
(the first supremum is over all decompositions of the state $\rho$
into countable convex combination of states of rank $\leq k$ while
the second one is over all probability measures with the barycenter
$\rho$ supported by states of rank $\leq k$).

For each $k$ the function $H_{k}$ may be called the \textit{entropy
approximator of order $k$} or briefly \textit{$k$-approximator}. By
construction the von Neumann entropy coincides with its
$k$-approximator on the set $\mathfrak{S}_{k}(\mathcal{H})$ of all
states of rank $\leq k$. For arbitrary state
$\rho\in\mathfrak{S}(\mathcal{H})$ the difference
$\Delta_{k}(\rho)=H(\rho)-H_{k}(\rho)$ between the von Neumann
entropy and its $k$-approximator can be expressed as follows
$$
\Delta_{k}(\rho)=\inf_{\{\pi_{i},\rho_{i}\}\in\mathcal{P}^{\mathrm{a}}_{\{\rho\}}(\mathfrak{S}_{k}(\mathcal{H}))}
\sum_{i}\pi_{i}H(\rho_{i}\|\rho)=\inf_{\mu\in\mathcal{P}_{\{\rho\}}(\mathfrak{S}_{k}(\mathcal{H}))}
\int_{\mathfrak{S}_{k}(\mathcal{H})}H(\sigma\|\rho)\mu(d\sigma),
$$
where $H(\cdot\|\cdot)$ is the relative entropy \cite{O&P,W} (the
first equality follows from expression (\ref{S-H}) below, the second
one -- from proposition 1 in \cite{H-Sh-2}). The possibility to
express the value $\Delta_{k}(\rho)$ via the relative entropy is
essentially used in what follows (see lemma \ref{Delta} below).

The representation of the von Neumann entropy as a limit of the
increasing sequence  $\{H_{k}\}$ of concave continuous bounded
unitary invariant functions can be used for different purposes, in
particular, for construction of the increasing sequence of
continuous entanglement monotones providing approximation of the
Entanglement of Formation (see section 6 in \cite{Sh-9}). By
corollary \ref{a-problem+} this representation can be used for
proving continuity of the von Neumann entropy on a subset of states
by showing uniform convergence to zero of the sequence
$\{\Delta_{k}\}$ on this subset. The last property of a subset of
states, in what follows called the \textit{uniform approximation
property}, is considered in detail in the next subsection (in the
extended context of subsets of the positive cone of
trace\nobreakdash-\hspace{0pt}class operators).

\subsection{The uniform approximation property}

Since in many applications it is necessary to deal with the
following extensions (cf.\cite{L-2})
$$
S(A)=-\mathrm{Tr}A\log A\quad \textrm{and}\quad
H(A)=S(A)-\eta(\mathrm{Tr}A)
$$
of the von Neumann entropy to the cone
$\mathfrak{T}_{+}(\mathcal{H})$ of all positive
trace\nobreakdash-\hspace{0pt}class operators (where $\eta(x)=-x\log
x$), we will obtain the continuity conditions for the function
$A\mapsto H(A)$ on this extended domain.

In what follows the function $A\mapsto H(A)$ on the cone
$\mathfrak{T}_{+}(\mathcal{H})$ is called the \textit{quantum
entropy} while the function $\{x_i\}\mapsto
H(\{x_i\})=\sum_{i}\eta(x_{i})-\eta\left(\sum_{i}x_{i}\right)$ on
the positive cone of the space $l_{1}$, coinciding with the Shannon
entropy on the set $\mathfrak{P}_{+\infty}$ of probability
distributions, is called the \textit{classical entropy}.

The von Neumann entropy has the important property expressed in the
following inequality
\begin{equation}\label{w-k-ineq}
H\left(\sum_{i=1}^{n}\lambda_{i}\rho_{i}\right)\leq
\sum_{i=1}^{n}\lambda_{i}H(\rho_{i})+\sum_{i=1}^{n}\eta(\lambda_{i}),
\end{equation}
valid for arbitrary set $\{\rho_{i}\}_{i=1}^{n}$ of states and
probability distribution $\{\lambda_{i}\}_{i=1}^{n}$, where
$n\leq+\infty$ (proposition 6.2 in \cite{O&P} and the simple
approximation).

The definition and inequality (\ref{w-k-ineq}) with $n=2$ imply the
following properties of the quantum entropy
\begin{eqnarray}
H(\lambda A)=\lambda H(A),\quad\quad\quad\quad\quad\quad\quad\quad\quad\quad\quad \label{H-fun-eq}\\
H(A)+H(B-A)\leq H(B)\leq H(A)+H(B-A)+\mathrm{Tr}B
h_{2}\left(\frac{\mathrm{Tr}A}{\mathrm{Tr}B}\right)\label{H-fun-ineq},
\end{eqnarray}
where $A,B\in\mathfrak{T}_{+}(\mathcal{H}),\; A\leq B,$ $\lambda\geq
0\,$ and $\,h_{2}(x)=\eta(x)+\eta(1-x)$.

Note that
\begin{equation}\label{S-H}
S(A)-\sum_{i}\pi_{i}S(A_{i})=\sum_{i}\pi_{i}H(A_{i}\|A)
\end{equation}
for arbitrary ensemble $\{\pi_{i}, A_{i}\}$ of operators in
$\mathfrak{T}_{+}(\mathcal{H})$ with the average $A$, where
$H(\cdot\|\cdot)$ is the (extended) relative entropy defined for
arbitrary operators $A$ and $B$ in $\mathfrak{T}_{+}(\mathcal{H})$
as follows (cf.\cite{L-2})
$$
H(A\,\|B)=\sum_{i}\langle i|\,(A\log A-A\log B+B-A)\,|i\rangle,
$$
where $\{|i\rangle\}$ is the orthonormal basis of eigenvectors of
$A$ and it is assumed that $H(A\,\|B)=+\infty$ if $\,\mathrm{supp}A$
is not contained in $\mathrm{supp}B$. It is easy to verify that
\begin{equation}\label{m-p-r-e}
H(\lambda A\,\|\lambda B)=\lambda H(A\,\|B),\quad \lambda\geq0.
\end{equation}

For given natural $k$ consider the function
$$
H_{k}(A)=\sup_{\{\pi_{i},A_{i}\}\in\mathcal{P}^{\mathrm{a}}_{\{A\}}(\mathfrak{T}_{+}^{k}(\mathcal{H}))}
\sum_{i}\pi_{i}H(A_{i})$$ on the set $\mathfrak{T}_{+}(\mathcal{H})$
(the supremum is over all decompositions of the operator $A$ into
countable convex combination of operators of rank $\leq k$). By
using (\ref{H-fun-eq}) it is easy to see that the restriction of the
above function $H_{k}$ to the set $\mathfrak{S}(\mathcal{H})$
coincides with the $k$-approximator of the von Neumann entropy
defined in the first part of this section (so, we use the same
notation) and that
$$
H_{k}(\lambda A)=\lambda H_{k}(A),\quad
A\in\mathfrak{T}_{+}(\mathcal{H}),\;\lambda\geq0.
$$
Thus we have
\begin{equation}\label{coincidence}
H_{k}(A)=\|A\|_{1}\hat{H}^{\sigma}_{k}(\|A\|_{1}^{-1}A)\leq
\|A\|_{1}\log k,\quad A\in\mathfrak{T}_{+}(\mathcal{H}).
\end{equation}

The contribution of the strong stability property of the set
$\mathfrak{S}(\mathcal{H})$ to the below results is based on the
following observation.\vspace{5pt}

\begin{lemma}\label{ssp-contribution}
\textit{For arbitrary natural $k$ the function $A\mapsto H_{k}(A)$
is continuous on the cone $\,\mathfrak{T}_{+}(\mathcal{H})$.
}\end{lemma}\vspace{5pt}

\textbf{Proof.} By means of (\ref{coincidence}) the assertion of the
lemma follows from corollary \ref{k-approx-c} showing continuity of
the function $\rho\mapsto \hat{H}^{\sigma}_{k}(\rho)$ on the set
$\mathfrak{S}(\mathcal{H})$. $\square$

For given natural $k$ consider the function
\begin{equation}\label{Delta-def}
\Delta_{k}(A)=\inf_{\{\pi_{i},A_{i}\}\in\mathcal{P}^{\mathrm{a}}_{\{A\}}(\mathfrak{T}_{+}^{k}(\mathcal{H}))}
\sum_{i}\pi_{i}H(A_{i}\|A)
\end{equation}
on the set $\mathfrak{T}_{+}(\mathcal{H})$ (the infimum is over all
decompositions of the operator $A$ into countable convex combination
of operators of rank $\leq k$).

It follows from (\ref{m-p-r-e}) that
\begin{equation}\label{Delta-m}
\Delta_{k}(\lambda A)=\lambda \Delta_{k}(A),\quad
A\in\mathfrak{T}_{+}(\mathcal{H}),\;\lambda\geq0.
\end{equation}

By lemma \ref{Delta} below the restriction of the function
$\Delta_{k}$ defined in (\ref{Delta-def}) to the set
$\mathfrak{S}(\mathcal{H})$ coincides with the function
$\Delta_{k}=H-H_{k}$ defined in the first part of this section (so,
we use the same notation).\vspace{5pt}

We will use the following properties of the function
$\Delta_{k}$.\vspace{5pt}

\begin{lemma}\label{Delta}
\textit{For each natural $k$ the following assertions hold:}
\begin{enumerate}[A)]
    \item \textit{For an arbitrary operator
$A\in\mathfrak{T}_{+}(\mathcal{H})$ the infimum in definition
(\ref{Delta-def}) of the value $\Delta_{k}(A)$ can be taken over the
subset of
$\,\mathcal{P}^{\mathrm{a}}_{\{A\}}(\mathfrak{T}_{+}^{k}(\mathcal{H}))$
consisting of ensembles $\{\pi_{i},A_{i}\}$ such that
$\,\mathrm{Tr}A_{i}=\mathrm{Tr}A$ for all $\,i$ and hence}
$$
\Delta_{k}(A)=H(A)-H_{k}(A).
$$
\item \textit{The function $\,\mathfrak{T}_{+}(\mathcal{H})\ni A\mapsto\Delta_{k}(A)$ is nonnegative lower
semicontinuous unitary invariant  and homogenous in the sense of
(\ref{Delta-m}).
$\Delta_{k}^{-1}(0)=\mathfrak{T}^{k}_{+}(\mathcal{H})$. Continuity
of this function on a subset
$\mathcal{A}\subset\mathfrak{T}_{+}(\mathcal{H})$ means continuity
of the quantum entropy on the subset $\mathcal{A}$.}

\item \textit{The function $A\mapsto\Delta_{k}(A)$ is monotone with
respect to the operator order:}
$$
A\leq B \quad\Rightarrow\quad\Delta_{k}(A)\leq \Delta_{k}(B),\qquad
\forall A,\,B\in\mathfrak{T}_{+}(\mathcal{H}).
$$

\item \textit{Let $\{\lambda_{i}(A)\}$ be the sequence of the eigenvalues of the operator
$A\in\mathfrak{T}_{+}(\mathcal{H})$ arranged in nonincreasing
order\footnote{It is possible to take the sequence
$\{\lambda_{i}(A)\}$ in arbitrary order but the corresponding
sequence $\{\lambda^{k}_{i}(A)\}$ is most close to the sequence
$(\|A\|_{1},0,0,...)$ having zero entropy provided that the
nonincreasing order is used. The relation between $\Delta_{k}(A)$
and $\widetilde{\Delta}_{k}(A)$ is considered in remark
\ref{noncoincidence} below.} then
$$
\Delta_{k}(A)\leq \widetilde{\Delta}_{k}(A)\doteq
H(\{\lambda_{i}^{k}(A)\})=\sum_{i=1}^{+\infty}\eta(\lambda_{i}^{k}(A))-\eta\left(\|A\|_{1}\right),
$$
where the sequence $\{\lambda_{i}^{k}(A)\}$ is the $k$-order
coarse-graining of the sequence $\{\lambda_{i}(A)\}$, t.i.
$\,\lambda_{i}^{k}(A)=\lambda_{(i-1)k+1}(A)+...+\lambda_{ik}(A)\,$
for all $\,i=1,2,...$}

\item \textit{For
arbitrary operators $A$ in $\mathfrak{T}_{+}(\mathcal{H})$ and $C$
in $\mathfrak{B}(\mathcal{H})$ the following inequality holds}
$$
\Delta_{k}(CAC^{*})\leq \|C\|^{2}\Delta_{k}(A).
$$

\item\textit{For an arbitrary operator $A$ in $\mathfrak{T}_{+}(\mathcal{H})$ and
an arbitrary sequence $\,\{P_{n}\}$ of projectors in
$\,\mathfrak{B}(\mathcal{H})$ strongly converging to the identity
operator $I_{\mathcal{H}}$ the following relation holds}
$$
\lim_{n\rightarrow+\infty}\Delta_{k}(P_{n}AP_{n})=\Delta_{k}(A).
$$

\item \textit{For an arbitrary operator $A$ in $\mathfrak{T}_{+}(\mathcal{H})$ and an arbitrary family $\,\{P_{i}\}_{i=1}^{m}$
of mutually orthogonal projectors in $\,\mathfrak{B}(\mathcal{H})$
($m\leq+\infty$) the following inequality holds}
$$
\Delta_{k}(A)\geq \sum_{i=1}^{m}\Delta_{k}(P_{i}AP_{i}).
$$

\item \textit{For an arbitrary operator $A$ in
$\mathfrak{T}_{+}(\mathcal{H})$ and an arbitrary quantum operation
$\Phi:\mathfrak{T}(\mathcal{H})\rightarrow\mathfrak{T}(\mathcal{H})$
having the Kraus representation consisting of $\,\leq n$ summands
the following inequality holds}
$$
\Delta_{nk}(\Phi(A))\leq \Delta_{k}(A).
$$

\item \textit{For an arbitrary finite set $\,\{A_{i}\}_{i=1}^{m}$ of
operators in $\,\mathfrak{T}_{+}(\mathcal{H})$ and corresponding set
$\,\{k_{i}\}_{i=1}^{m}$ of natural numbers the following inequality
holds}
$$
\Delta_{k_{1}+k_{2}+...+k_{m}}\left(\sum_{i=1}^{m}A_{i}\right)\leq
\sum_{i=1}^{m}\Delta_{k_{i}}(A_{i}).
$$

\item \textit{For an arbitrary countable set  $\,\{A_{i}\}_{i=1}^{+\infty}$ of
operators in $\mathfrak{T}_{+}(\mathcal{H})$, probability
distribution $\{\lambda_{i}\}_{i=1}^{+\infty}$ and natural $\,m$ the
following inequality holds}
$$
\begin{array}{c}
\displaystyle\Delta_{mk}\left(\sum_{i=1}^{+\infty}\lambda_{i}A_{i}\right)\leq
\sum_{i=1}^{+\infty}\lambda_{i}\Delta_{k}(A_{i})+\sum_{i=m}^{+\infty}\lambda_{i}H\left(A_{i}\|
\left(\sum_{i=m}^{+\infty}\lambda_{i}\right)^{-1}\sum_{i=m}^{+\infty}\lambda_{i}A_{i}\right)\\\\
\displaystyle\leq\sum_{i=1}^{+\infty}\lambda_{i}\Delta_{k}(A_{i})+\sup_{i\geq
m}\|A_{i}\|_{1}H\left(\{\lambda_{i}\}_{i\geq m}\right).
\end{array}
$$
\end{enumerate}
\end{lemma}
\textbf{Proof.} A) For arbitrary ensemble $\{\pi_{i}, A_{i}\}$ in
$\mathcal{P}^{\mathrm{a}}_{\{A\}}(\mathfrak{T}_{+}^{k}(\mathcal{H}))$
one can consider ensemble $\{\lambda_{i}, B_{i}\}$ in
$\mathcal{P}^{\mathrm{a}}_{\{A\}}(\mathfrak{T}_{+}^{k}(\mathcal{H}))$,
where $\lambda_{i}=\pi_{i}\|A_{i}\|_{1}\|A\|_{1}^{-1}$ and
$B_{i}=A_{i}\|A\|_{1}\|A_{i}\|_{1}^{-1}$, such that
$$
\begin{array}{c}
\displaystyle\sum_{i}\lambda_{i}H(B_{i}\|A)=\sum_{i}\pi_{i}H(A_{i}\|A)\\\displaystyle-
\left(\eta\left(\|A\|_{1}\right)-
\sum_{i}\pi_{i}\eta\left(\|A_{i}\|_{1}\right)\right)\leq\sum_{i}\pi_{i}H(A_{i}\|A),
\end{array}
$$
where the last inequality follows from concavity of the function
$\eta$, since $\sum_{i}\pi_{i}\|A_{i}\|_{1}=\|A\|_{1}$. By
(\ref{H-fun-eq}) and (\ref{coincidence}) this implies
$\Delta_{k}(A)=H(A)-H_{k}(A)$.

B) Lemma \ref{ssp-contribution} and assertion A imply the first and
the third parts of this assertion. To prove the second one note that
the inclusion
$\mathfrak{T}^{k}_{+}(\mathcal{H})\subseteq\Delta_{k}^{-1}(0)$
follows from the definition of the function $\Delta_{k}$ while the
converse inclusion is easily derived from the implication
$\rho\in\mathfrak{S}(\mathcal{H})\setminus\mathfrak{S}_{k}(\mathcal{H})\,\Rightarrow\,
H(\rho)>H_{k}(\rho)$, which follows from strict concavity of the von
Neumann entropy and the last assertion of proposition
\ref{k-approx-us}, implying attainability of the supremum in the
second (continuous) expression in the definition of the function
$H_{k}(\rho)$.

C) If $A\leq B$ then there exists contraction $C$ such that
$A=CBC^{*}$. Indeed, on the subspace $\mathrm{supp}B$ this
contraction is constructed as the continuous extension to this
subspace of the linear operator $A^{1/2}B^{-1/2}$ defined on the
linear hull of the eigenvectors of the operator $B$ corresponding to
the positive eigenvalues while on the subspace
$\mathcal{H}\ominus\mathrm{supp}B$ it acts as the zero operator.
Hence this assertion follows from assertion H proved below.

D) Let $\{P_{i}^{k}\}_{i}$ be the sequence of spectral projectors of
the operator $A$ such that the projector $P^{k}_{i}$ corresponds to
the eigenvalues $\lambda_{(i-1)k+1}(A),...,\lambda_{ik}(A)$. Then
$\lambda_{i}^{k}(A)=\mathrm{Tr}P^{k}_{i}A$ for all $i$ and the
ensemble $\{\pi^{k}_{i}, (\pi^{k}_{i})^{-1}P^{k}_{i}A\}$, where
$\pi^{k}_{i}=\lambda_{i}^{k}(A)\|A\|_{1}^{-1}$, belongs to the set
$\mathcal{P}^{\mathrm{a}}_{\{A\}}(\mathfrak{T}_{+}^{k}(\mathcal{H}))$.
Hence
$$
\Delta_{k}(A)\leq\sum_{i}\pi^{k}_{i}H((\pi^{k}_{i})^{-1}P^{k}_{i}A\|A)=H(\{\lambda_{i}^{k}(A)\}).
$$

E) By means of (\ref{Delta-m}) this follows from assertion H proved
below.

F) By lower semicontinuity of the function $\Delta_{k}$ (assertion
B) this follows from assertion E.

G) It is sufficient to prove that
$$
\Delta_{k}(A)\geq\Delta_{k}(PAP)+\Delta_{k}(\bar{P}A\bar{P}),
$$
where $\bar{P}=I_{\mathcal{H}}-P$, for arbitrary projector $P$. This
inequality is easily proved by using the definition of the function
$\Delta_{k}$ and the inequality
$$
H(A\|B)\geq H(PAP\|PBP)+H(\bar{P}A\bar{P}\|\bar{P}B\bar{P})
$$
valid for arbitrary operators $A$ and $B$ in
$\mathfrak{T}_{+}(\mathcal{H})$ (lemma 3 in \cite{L-2}).

H) This follows from monotonicity of the relative entropy since for
an arbitrary ensemble $\{\pi_{i}, A_{i}\}$ in
$\mathcal{P}^{\mathrm{a}}_{\{A\}}(\mathfrak{T}_{+}^{k}(\mathcal{H}))$
the ensemble $\{\pi_{i}, \Phi(A_{i})\}$ lies in
$\mathcal{P}^{\mathrm{a}}_{\{\Phi(A)\}}(\mathfrak{T}_{+}^{nk}(\mathcal{H}))$.

I) By means of (\ref{Delta-m}) it is sufficient to show that
\begin{equation}\label{ppp}
\Delta_{k'+k''}(\gamma A+(1-\gamma)B)\leq
\gamma\Delta_{k'}(A)+(1-\gamma)\Delta_{k''}(B)
\end{equation}
for arbitrary operators $A$ and $B$  in
$\mathfrak{T}_{+}(\mathcal{H})$ and $\gamma\in[0,1]$. For given $k'$
and $k''$ let $\{\pi_{i}, A_{i}\}_{i}$ and $\{\lambda_{j},
B_{j}\}_{j}$ be ensembles of operators of rank $\leq k'$ with the
average $A$  and of rank $\leq k''$ with the average $B$
correspondingly. Then the ensemble $\{\pi_{i}\lambda_{j}, \gamma
A_{i}+(1-\gamma)B_{j}\}_{i,j}$ has the average $\gamma
A+(1-\gamma)B$ and consists of operators of rank $\leq k'+k''$. By
joint convexity of the relative entropy we have
$$
\begin{array}{c}
\displaystyle\Delta_{k'+k''}(\gamma A+(1-\gamma) B)\leq
\sum_{i,j}\pi_{i}\lambda_{j}H(\gamma A_{i}+(1-\gamma)B_{j}\|\gamma
A+(1-\gamma)B)\\\displaystyle\leq \gamma\sum_{i}\pi_{i}H(A_{i}\|
A)+(1-\gamma)\sum_{j}\lambda_{j}H(B_{j}\|B),
\end{array}
$$
which implies inequality (\ref{ppp}).

J) The first inequality with $m=1$ is easily derived from the
definition of the function $\Delta_{k}$ by using Donald's identity
$$
\sum_{i}\pi_{i}H(A_{i}\|B)=\sum_{i}\pi_{i}H(A_{i}\|A)+H(A\|B)
$$
valid for arbitrary ensemble $\{\pi_{i},A_{i}\}$ of positive
trace\nobreakdash-\hspace{0pt}class operators with the average $A$
and arbitrary trace\nobreakdash-\hspace{0pt}class operator $B$
\cite{O&P}. The case $m>1$ is reduced to the case $m=1$  by applying
assertion I (with (\ref{Delta-m})) to the sum
$\sum_{i=1}^{m}A'_{i}$, where $A'_{i}=\lambda_{i}A_{i}$ for
$i=\overline{1,m-1}$ and
$A'_{m}=\sum_{i=m}^{+\infty}\lambda_{i}A_{i}$.

The second inequality follows from the estimation
$$
\sum_{i}\pi_{i}H(A_{i}\|A)\leq
\sup_{i}\|A_{i}\|_{1}H\left(\{\pi_{i}\}\right)
$$
valid for arbitrary ensemble $\{\pi_{i}, A_{i}\}$ of
trace\nobreakdash-\hspace{0pt}class operators with the average $A$,
which can be proved by using monotonicity of the relative entropy:
$$
\sum_{i}\pi_{i}H(A_{i}\|A)=c\sum_{i}\pi_{i}H(\Phi(\rho_{i})\|\Phi(\rho))\leq
c\sum_{i}\pi_{i}H(\rho_{i}\|\rho)=cH\left(\{\pi_{i}\}\right),
$$
where $c=\sup_{i}\|A_{i}\|_{1}$, $\Phi(\cdot)=c^{-1}\sum_{i}\langle
i|\cdot|i\rangle A_{i}$, $\rho_{i}=|i\rangle\langle i|$ and
$\rho=\sum_{i}\pi_{i}\rho_{i}$. $\square$ \vspace{5pt}

\begin{remark}\label{noncoincidence} It is easy to show that the upper bound
$\widetilde{\Delta}_{k}(A)$ in assertion D of lemma \ref{Delta}
obtained by using the spectral decomposition of the operator $A$
tends to zero if $H(A)<+\infty$, which provides the additional proof
of convergence of the sequence $\{H_{k}\}$ to the function $H$ on
the cone $\mathfrak{T}_{+}(\mathcal{H})$. Noncoincidence of the
functions $\widetilde{\Delta}_{k}$ and $\Delta_{k}$, t.i. existence
of such operator $A$ in $\mathfrak{T}_{+}(\mathcal{H})$ that
$\Delta_{k}(A)<\widetilde{\Delta}_{k}(A)$, can be shown by the
following example.

Let $\rho$ be the chaotic state in a particular $3\textrm{-D}$
subspace $\mathcal{H}_{0}\subset\mathcal{H}$. It is clear that
$\widetilde{\Delta}_{2}(\rho)=\log 3-\frac{2}{3}\log 2\approx 0.64$
(we use the natural logarithm).

In the subspace $\mathcal{H}_{0}$ consider four unit vectors
$$
|\varphi_{1}\rangle=\left[\begin{array}{c}
        \;1\; \\
        \;0\; \\
        \;0\; \\
        \end{array}\right],
|\varphi_{2}\rangle=\left[\begin{array}{c}
        \;-1/2\; \\
        \;\sqrt{3}/2\; \\
        \;0\; \\
        \end{array}\right],
|\varphi_{3}\rangle=\left[\begin{array}{c}
        \;-1/2\; \\
        \;-\sqrt{3}/2\; \\
        \;0\; \\
        \end{array}\right],
|\varphi_{4}\rangle=\left[\begin{array}{c}
        \;0\; \\
        \;0\; \\
        \;1\; \\
        \end{array}\right].
$$
By direct calculation of eigenvalues one can show that the two rank
states
$\rho_{1}=\frac{1}{2}|\varphi_{1}\rangle\langle\varphi_{1}|+\frac{1}{2}|\varphi_{2}\rangle\langle\varphi_{2}|$
and
$\rho_{2}=\frac{2}{5}|\varphi_{3}\rangle\langle\varphi_{3}|+\frac{3}{5}|\varphi_{4}\rangle\langle\varphi_{4}|$
have the entropies $H(\rho_{1})\approx 0.57$ and $H(\rho_{2})\approx
0.67$. Since $\frac{4}{9}\rho_{1}+\frac{5}{9}\rho_{2}=\rho$ we can
conclude that $H_{2}(\rho)\geq
\frac{4}{9}H(\rho_{1})+\frac{5}{9}H(\rho_{2})\approx 0.63$. Thus
$\Delta_{2}(\rho)=H(\rho)-H_{2}(\rho)<\widetilde{\Delta}_{2}(\rho)$.
$\square$
\end{remark}\vspace{5pt}

The following notion plays the central role in this
paper.\vspace{5pt}

\begin{definition}\label{UA-property}
\textit{A subset $\mathcal{A}$ of $\,\mathfrak{T}_{+}(\mathcal{H})$
has the uniform approximation property (briefly the
UA\nobreakdash-\hspace{0pt}property) if}
$$
\lim_{k\rightarrow+\infty}\sup_{A\in\mathcal{A}}\Delta_{k}(A)=0.
$$
\end{definition}\vspace{5pt}

Importance of the UA\nobreakdash-\hspace{0pt}property is justified
by its close relation to continuity of the quantum entropy
considered in theorem \ref{cont-cond} in the next subsection.
Usefulness of this relation is based on the following observation,
showing the conservation of the UA\nobreakdash-\hspace{0pt}property
under different set-operations.\pagebreak

\begin{property}\label{UA-property-c} \textit{Let $\mathcal{A}$ be a subset
of $\;\mathfrak{T}_{+}(\mathcal{H})$ having the
UA\nobreakdash-\hspace{0pt}property.}\vspace{5pt}

\begin{enumerate}[A)]

\item \textit{The UA\nobreakdash-\hspace{0pt}property holds for the
closure $\mathrm{cl}(\mathcal{A})$ of the set
$\mathcal{A}$.}\vspace{5pt}

\item \textit{For each $\lambda>0$ the
UA\nobreakdash-\hspace{0pt}property holds for the set
$$
M_{\lambda}(\mathcal{A})=\{\lambda A\,| A\in\mathcal{A}\}.
$$}

\item \textit{If $\,\inf_{A\in\mathcal{A}}\|A\|_{1}>0$ then the
UA\nobreakdash-\hspace{0pt}property holds for the set
$$
E(\mathcal{A})=\left\{\lambda A\,|\,A\in\mathcal{A}, \lambda\geq
0\right\}\cap\mathfrak{T}_{1}(\mathcal{H}).
$$}

\item \textit{For each natural $m$ the UA\nobreakdash-\hspace{0pt}property holds for the
set}
$$
\mathrm{co}_{m}(\mathcal{A})=\left\{\left.\sum_{i=1}^{m}
\pi_{i}A_{i}\,\right|\, \{\pi_{i}\}\in\mathfrak{P}_{m},
\{A_{i}\}\subseteq\mathcal{A}\right\}.
$$
\textit{If the set $\mathcal{A}$ is bounded then the
UA\nobreakdash-\hspace{0pt}property holds for the set
$$
\mathrm{co}_{\mathfrak{P}}(\mathcal{A})=\left\{\left.\sum_{i=1}^{+\infty}
\pi_{i}A_{i}\,\right|\, \{\pi_{i}\}\in\mathfrak{P},
\{A_{i}\}\subseteq\mathcal{A}\right\},
$$
where $\,\mathfrak{P}$ is a subset of $\,\mathfrak{P}_{+\infty}$
such that
$\displaystyle\lim_{m\rightarrow+\infty}\sup_{\{\pi_{i}\}\in\mathfrak{P}}H(\{\pi_{i}\}_{i>m})=0$.}

\item \textit{The UA\nobreakdash-\hspace{0pt}property holds for the
sets $$
D(\mathcal{A})=\left\{\left.B\in\mathfrak{T}_{+}(\mathcal{H})\,\right|\,\exists
A\in\mathcal{A}:B\leq A\right\} $$
and
$$
\widetilde{D}(\mathcal{A})=\left\{\left.B\in\mathfrak{T}_{+}(\mathcal{H})\,\right|\,\exists
A\in\mathcal{A}:B\lessdot A\right\},
$$
where $B\lessdot A$ means that the sequence $\{\lambda_{i}(B)\}$ of
eigenvalues of the operator $B$ is majorized by the sequence
$\{\lambda_{i}(A)\}$ of eigenvalues of the operator $A$ in the sense
$\lambda_{i}(B)\leq\lambda_{i}(A)$ for all $i$;} \vspace{5pt}

\textit{If the set $\mathcal{A}$ is compact and does not contain the
null operator then the UA\nobreakdash-\hspace{0pt}property holds for
the set
$$
\widehat{D}(\mathcal{A})=\left\{\left.B\in\mathfrak{T}_{1}(\mathcal{H})\,\right|\,\exists
A\in\mathcal{A}:B\|B\|_{1}^{-1}\prec A\|A\|_{1}^{-1}\right\}
$$
where $\rho\prec\sigma$ means that the state $\sigma$ is more
chaotic than the state $\rho$ in the Uhlmann sense \cite{U+,W+},
t.i. for the sequences $\{\lambda_{i}(\rho)\}$ and
$\{\lambda_{i}(\sigma)\}$ of eigenvalues of the states $\rho$ and
$\sigma$ arranged in nonicreasing order the inequality
$\sum_{i=1}^{n}\lambda_{i}(\rho)\geq\sum_{i=1}^{n}\lambda_{i}(\sigma)$
holds for each natural $n$.}

\item \textit{The
UA\nobreakdash-\hspace{0pt}property holds for the
sets\footnote{$\mathfrak{F}_{\leq 1}^{n}(\mathcal{H})$ is the set of
all quantum operations having the Kraus representation consisting of
$\,\leq n$ summands (see section 2).}
$$
Q_{n}(\mathcal{A})=\left\{\displaystyle
\Phi(A)\,\left|\,\Phi\in\mathfrak{F}_{\leq 1}^{n}(\mathcal{H}),
A\in\mathcal{A}\right.\right\}, \quad n\in\mathbb{N},
$$}
\textit{and
$$
Q_{\mathfrak{F}}(\mathcal{A})=\left\{\displaystyle
\Phi(A)\,\left|\,\Phi\in\mathfrak{F},
A\in\mathcal{A}\right.\right\},
$$
where $\,\mathfrak{F}$ is a subset of $\,\mathfrak{F}_{\leq
1}(\mathcal{H})$ such that for the corresponding set $\mathfrak{V}$
of sequences $\{V_{j}\}_{j=1}^{+\infty}$ of Kraus operators the
following two conditions holds:}\footnote{The ways to show validity
of these conditions are considered in corollary \ref{new} below.}

\begin{enumerate}[1)]
    \item \textit{either  $\,\mathrm{Ran}V^{*}_{j}\perp\mathrm{Ran}V^{*}_{j'}$ for
    all $\,\{V_{j}\}_{j=1}^{+\infty}\in\mathfrak{V}$ and all
    $j\neq j'$ exceeding some natural $n$ or $\;\lim_{m\rightarrow+\infty}\sup_{\{V_{j}\}\in\mathfrak{V},A\in\mathcal{A}}\sum_{j\geq
m}H\left(V_{j}AV_{j}^{*}\right)=0$;}
    \item
    $$
    \lim_{m\rightarrow+\infty}\sup_{\{V_{j}\}\in\mathfrak{V},A\in\mathcal{A}}H\left(\{\mathrm{Tr}V_{j}AV_{j}^{*}\}_{j\geq
    m}\right)=0.
    $$
\end{enumerate}

\end{enumerate}

\end{property}\vspace{5pt}

\begin{remark}\label{UA-property-c-r}
In connection with assertion D one can note that the
UA\nobreakdash-\hspace{0pt}property of a set $\mathcal{A}$ does not
imply the UA\nobreakdash-\hspace{0pt}property its
$\sigma$\nobreakdash-\hspace{0pt}convex hull
$\sigma\textrm{-}\mathrm{co}(\mathcal{A})=\left\{\sum_{i=1}^{+\infty}
\pi_{i}A_{i}\,|\, \{\pi_{i}\}\in\mathfrak{P}_{+\infty},
\{A_{i}\}\subseteq\mathcal{A}\right\}$ even if the set $\mathcal{A}$
is compact. As an example one can consider the converging sequence
of pure states from the example in section 5.1 in \cite{Sh-4}, such
that the von Neumann entropy is not continuous on the
$\sigma$\nobreakdash-\hspace{0pt}convex hull of this sequence (since
the UA\nobreakdash-\hspace{0pt}property implies continuity of the
entropy by lemma \ref{ssp-contribution}).

Note that the condition
$\lim_{m\rightarrow+\infty}\sup_{\{\pi_{i}\}\in\mathfrak{P}}H(\{\pi_{i}\}_{i>m})=0$
means continuity of the classical entropy on the set $\mathfrak{P}$
provided that this set is compact.
\end{remark}\vspace{5pt}

\textbf{Proof of proposition \ref{UA-property-c}.} A) This follows
from lower semicontinuity of the function $\Delta_{k}$ on the set
$\mathfrak{T}_{+}(\mathcal{H})$ for each $k$ (lemma \ref{Delta}B).

B) This is an obvious corollary of (\ref{Delta-m}).

C) This also follows from (\ref{Delta-m}) since
$$
\sup_{B\in E(\mathcal{A})} \{\lambda\,|\,B=\lambda A,
A\in\mathcal{A}\}\leq
\left(\inf_{A\in\mathcal{A}}\|A\|_{1}\right)^{-1}.
$$

D) The first part follows from lemma \ref{Delta}I and
(\ref{Delta-m}) implying
$$
\Delta_{km}\left(\sum_{i=1}^{m}\pi_{i}A_{i}\right)\leq
\sum_{i=1}^{m}\pi_{i}\Delta_{k}(A_{i}),\quad
\forall\,\{\pi_{i}\}_{i=1}^{m}\in\mathfrak{P}_{m}.
$$

The second part follows from lemma \ref{Delta}J since for arbitrary
$k$ and $m$ it implies
$$
\begin{array}{c}
\displaystyle\Delta_{km}\left(\sum_{i=1}^{+\infty}\pi_{i}A_{i}\right)\leq
\sum_{i=1}^{+\infty}\pi_{i}\Delta_{k}(A_{i})+\sup_{i\geq
m}\|A_{i}\|_{1}H\left(\{\pi_{i}\}_{i\geq
m}\right)\\\\\displaystyle\leq
\sup_{A\in\mathcal{A}}\Delta_{k}(A)+\sup_{A\in\mathcal{A}}\|A\|_{1}H\left(\{\pi_{i}\}_{i\geq
m}\right),\quad \forall\,
\{A_{i}\}_{i=1}^{+\infty}\subseteq\mathcal{A},\;\forall\,
\{\pi_{i}\}_{i=1}^{+\infty}\in\mathfrak{P}_{+\infty}.
\end{array}
$$

E) The first part follows from lemma \ref{Delta}C and unitary
invariance of the function $\Delta_{k}$.

The second part follows from lemma \ref{Delta}D and lemma
\ref{s-delta} below since
$$
B\|B\|_{1}^{-1}\prec A\|A\|_{1}^{-1}\quad\Rightarrow\quad
H(\{\lambda_{i}^{k}(B)\})\leq \|B\|_{1}\|A\|_{1}^{-1}
H(\{\lambda_{i}^{k}(A)\})
$$
for each natural $k$ by Shur concavity of the von Neumann entropy
\cite{W+}.

F) The first part follows from lemma \ref{Delta}H.

To prove the second part note that lemma \ref{Delta}J implies the
inequality
$$
\Delta_{km}\left(\sum_{j=1}^{+\infty}V_{j}AV_{j}^{*}\right)\leq
\sum_{j=1}^{+\infty}\Delta_{k}(V_{j}AV_{j}^{*})+H\left(\{\mathrm{Tr}V_{j}AV_{j}^{*}\}_{j\geq
m}\right)
$$
Thus it is sufficient to show that condition 1) implies
\begin{equation}\label{uc}
\lim_{k\rightarrow+\infty}\sup_{\{V_{j}\}\in\mathfrak{V},A\in\mathcal{A}}\sum_{j=1}^{+\infty}\Delta_{k}(V_{j}AV_{j}^{*})=0.
\end{equation}

If the first alternative in condition 1) holds then assertions E and
G of lemma \ref{Delta} provide the estimation
$$
\begin{array}{c}
\displaystyle\sum_{j=1}^{+\infty}\Delta_{k}(V_{j}AV_{j}^{*})=\sum_{j=1}^{n}\Delta_{k}(V_{j}AV_{j}^{*})+
\sum_{j>n}\Delta_{k}(V_{j}AV_{j}^{*})
\\\\\displaystyle\leq
n\Delta_{k}(A)+\sum_{j>n}\Delta_{k}(P_{j}AP_{j})\leq
(n+1)\Delta_{k}(A),\quad \forall\,\{V_{j}\}\in\mathfrak{V},
\end{array}
$$
where $P_{j}$ is the projector on the subspace
$\mathrm{Ran}V^{*}_{j}$, which implies (\ref{uc}) by the
UA\nobreakdash-\hspace{0pt}property of the set $\mathcal{A}$.

If the second alternative in condition 1) holds then the similar
estimation, in which the term
$\sum_{j>n}\Delta_{k}(V_{j}AV_{j}^{*})$ is majorized by
$\sum_{j>n}H(V_{j}AV_{j}^{*})$ also implies (\ref{uc}) by the
UA\nobreakdash-\hspace{0pt}property of the set $\mathcal{A}$.
$\square$ \vspace{5pt}

\begin{lemma}\label{s-delta}
\textit{Let $\mathcal{A}$ be a compact subset of
$\,\mathfrak{T}_{+}(\mathcal{H})$ having the
UA\nobreakdash-\hspace{0pt}property. Then
$\,\lim_{k\rightarrow+\infty}\sup_{A\in\mathcal{A}}\widetilde{\Delta}_{k}(A)=0$,
where $\widetilde{\Delta}_{k}$ is the upper bound for the function
$\Delta_{k}$ defined in lemma \ref{Delta}D.}
\end{lemma} \vspace{5pt}

\textbf{Proof.} By lemma \ref{ssp-contribution} the
UA\nobreakdash-\hspace{0pt}property of the set $\mathcal{A}$ implies
continuity of the function $A\mapsto H(A)$ on this set.

Let $\{P_{i}^{k}\}_{i}$ be the sequence of spectral projectors of
the operator $A$ defined in the proof of assertion D of lemma
\ref{Delta} and $\pi^{k}_{i}=\|A\|_{1}^{-1}\mathrm{Tr}P_{i}^{k}A$
for all $i$. By lemma 4 in \cite{L-2} the sequence of continuous
functions $A\mapsto H(P^{k}_{1}A)$ monotonously converges to the
function $A\mapsto H(A)$ as $k\rightarrow+\infty$. By Dini's lemma
this sequence converges uniformly on the set $\mathcal{A}$. This
implies the assertion of the lemma since
$$
\widetilde{\Delta}_{k}(A)=\sum_{i}\pi^{k}_{i}H((\pi^{k}_{i})^{-1}P^{k}_{i}A\|A)\leq
H(A)-H(P^{k}_{1}A),\quad A\in\mathcal{A} \quad\square.
$$

By definition the UA\nobreakdash-\hspace{0pt}property of sets
$\mathcal{A}$ and $\mathcal{B}$ implies the
UA\nobreakdash-\hspace{0pt}property of their union
$\mathcal{A}\cup\mathcal{B}$. By lemma 6I and proposition
\ref{UA-property-c}D we have the following observations.\vspace{5pt}

\begin{corollary}\label{UA-property-c-c}
\textit{Let $\mathcal{A}$ and $\mathcal{B}$ be subsets of
$\,\mathfrak{T}_{+}(\mathcal{H})$ having the
UA\nobreakdash-\hspace{0pt}property.}\vspace{5pt}

A) \textit{The UA\nobreakdash-\hspace{0pt}property holds for the
set\footnote{This set is called the Minkowski sum of the sets
$\mathcal{A}$ and $\mathcal{B}$;}
$\mathcal{A}\boxplus\mathcal{B}=\{A+B\,|\,A\in\mathcal{A},
B\in\mathcal{B}\}$;}\vspace{5pt}

B) \textit{The UA\nobreakdash-\hspace{0pt}property holds for the
convex closure $\overline{\mathrm{co}}(\mathcal{A}\cup\mathcal{B})$
of the union of $\mathcal{A}$ and $\mathcal{B}$ provided these sets
are convex.}
\end{corollary}
\vspace{5pt}

\subsection{The continuity conditions}

Lemmas \ref{ssp-contribution} and \ref{Delta}, Dini's lemma and
proposition \ref{UA-property-c} imply the following theorem,
containing the main results of this paper.\vspace{5pt}

\begin{theorem}\label{cont-cond}
A) \textit{If a set
$\mathcal{A}\subset\mathfrak{T}_{+}(\mathcal{H})$ has the
UA\nobreakdash-\hspace{0pt}property then the quantum entropy is
continuous on this set.}\vspace{5pt}

B) \textit{If the quantum entropy is continuous on a compact set
$\mathcal{A}\subset\mathfrak{T}_{+}(\mathcal{H})$ then this set has
the UA\nobreakdash-\hspace{0pt}property.}\vspace{5pt}

C) \textit{If a set
$\mathcal{A}\subset\mathfrak{T}_{+}(\mathcal{H})$ has the
UA\nobreakdash-\hspace{0pt}property then the quantum entropy is
continuous on the set $\Lambda(\mathcal{A})$, where $\Lambda$ is an
arbitrary finite composition of the
set\nobreakdash-\hspace{0pt}operations $\,\mathrm{cl}$,
$M_{\lambda}$, $E$, $\mathrm{co}_{m}$, $\mathrm{co}_{\mathfrak{P}}$,
$D$, $\widetilde{D}$, $\widehat{D}$, $Q_{n}$, $Q_{\mathfrak{F}}$
considered in proposition \ref{UA-property-c} with arbitrary
parameters $m,n\in\mathbb{N}$ and $\lambda>0$ provided the sets
$\,\mathfrak{P}$, $\mathfrak{F}$ and the arguments of $E$,
$\,\mathrm{co}_{\mathfrak{P}}$, $\widehat{D}$, $Q_{\mathfrak{F}}$
satisfy the conditions mentioned in this proposition.}
\end{theorem}\vspace{5pt}

\begin{remark}\label{ns-cont-cond-r}
As the simplest example showing importance of the compactness
condition in the second assertion of theorem \ref{cont-cond} one can
consider the set $\mathcal{A}=\{\lambda
\rho\,|\,\lambda\in\mathbb{R}_{+}\}$, where $\rho$ is an infinite
rank state with finite entropy.

The following example shows that the second assertion of theorem
\ref{cont-cond} can not be valid even for relatively compact convex
sets of states.

Let $\{\rho_{i}\}_{i\geq 0}$ be a sequence of finite rank states in
$\mathfrak{S}(\mathcal{H})$ such that $\rho_{0}$ is a pure state,
$H(\rho_{i})\geq1$ for all $i>0$,
$\mathrm{supp}\rho_{n}\subset\mathcal{H}\ominus\left(\bigoplus_{i=0}^{n-1}\mathrm{supp}\rho_{i}\right)$
and $\sum_{i=1}^{+\infty}e^{-\lambda H(\rho_{i})}<+\infty$ for all
$\lambda>0$. Let $\lambda_{i}=(H(\rho_{i}))^{-1}$ for each
$i\in\mathbb{N}$. Consider the sequence of states
$$
\sigma_{i}=(1-\lambda_{i})\rho_{0}+\lambda_{i}\rho_{i},\quad
i\in\mathbb{N},
$$
obviously converging to the state $\rho_{0}$.

In Appendix 7.2 it is proved that \textit{the von Neumann entropy is
continuous on the convex set
$\mathcal{A}=\sigma\textup{-}\mathrm{co}\left(\{\sigma_{i}\}_{i\in\mathbb{N}}\right)=
\left\{\sum_{i=1}^{+\infty}\pi_{i}\sigma_{i}\,|\,\{\pi_{i}\}\in\mathfrak{P}_{+\infty}\right\}$,
but it is not continuous on the set
$\,\mathrm{cl}\,(\mathcal{A})=\overline{\mathrm{co}}\left(\{\sigma_{i}\}_{i\in\mathbb{N}}\right)=\mathcal{A}\cup\{\rho_{0}\}$.}
By the first assertion of theorem \ref{cont-cond} and proposition
\ref{UA-property-c}A the UA\nobreakdash-\hspace{0pt}property does
not hold for the set $\mathcal{A}$.  $\square$
\end{remark}\vspace{5pt}

Show first that theorem \ref{cont-cond} makes possible to re-derive
the continuity conditions mentioned in the Introduction in the
generalized forms.\vspace{5pt}

\begin{example}\label{bounded-energy} Let $\{h_{i}\}$ be a nondecreasing sequence of nonnegative numbers
and $\,\mathfrak{P}_{\{h_{i}\},h}$ be the subset of
$\,\mathfrak{P}_{+\infty}$ consisting of probability distributions
$\{\pi_{i}\}$ satisfying the inequality $\sum_{i}h_{i}\pi_{i}\leq
h$. By lemma \ref{simple} in the Appendix the set
$\,\mathfrak{P}_{\{h_{i}\},h}$ satisfies the condition in
proposition \ref{UA-property-c}D if and only if
$\mathrm{g}\left(\{h_{i}\}\right)=\inf\left\{\lambda>0\,|\,\sum_{i}e^{-\lambda
h_{i}}<+\infty\right\}=0$. By theorem \ref{cont-cond}C the von
Neumann entropy is continuous on the set
$\,\mathrm{cl}(\mathrm{co}_{\mathfrak{P}_{\{h_{i}\},h}}(\mathfrak{S}_{k}(\mathcal{H})))$
for each $k$. This observation provides the another
proof\footnote{The original proof of this results is based on lower
semicontinuity of the function $\rho\mapsto
H(\rho\|\sigma_{\lambda})$, where
$\sigma_{\lambda}=(\mathrm{Tr}e^{-\lambda H})^{-1}e^{-\lambda H}$,
for all $\lambda>0$ \cite{O&P,W}.} of the well known result stated
that the entropy is continuous on the set
$\mathcal{K}_{H,h}=\{\rho\in\mathfrak{S}(\mathcal{H})\,|\,\mathrm{Tr}H\rho\leq
h\}$, where $H$ is an
$\mathfrak{H}$\nobreakdash-\hspace{0pt}operator such that
$\mathrm{g}(H)=\inf\left\{\lambda>0\,|\,\mathrm{Tr}e^{-\lambda
H}<+\infty\right\}=0$, since by using the extremal properties of
eigenvalues of a positive operator it is easy to see that the set
$\,\mathrm{cl}(\mathrm{co}_{\mathfrak{P}_{\{h_{i}\},h}}(\mathfrak{S}_{1}(\mathcal{H})))$,
where $\{h_{i}\}$ is the sequence of eigenvalues of the operator
$H$, contains the set $\mathcal{K}_{H,h}$ (and all its unitary
translations).

The von Neumann entropy is not continuous on the set
$\,\mathrm{cl}(\mathrm{co}_{\mathfrak{P}_{\{h_{i}\},h}}(\mathfrak{S}_{1}(\mathcal{H}))$
if $\mathrm{g}\left(\{h_{i}\}\right)>0$ since it is not continuous
on the set $\mathcal{K}_{H,h}$  if $\mathrm{g}(H)>0$ \cite{Sh-4}.
$\square$
\end{example}\vspace{5pt}

Theorem \ref{cont-cond} implies the following generalization of
Simon's dominated convergence theorems \cite{Simon}.\vspace{5pt}

\begin{corollary}\label{cont-cond++}
\textbf{(generalized Simon's convergence theorem)}\footnote{In the
original versions of these theorems the weaker topologies are used.
Since the set $D(\mathcal{A})$ is compact (by the compactness
criterion in proposition \ref{comp-crit} in the Appendix), the weak
operator topology on this set coincides with the trace norm
topology. The $\mu$\nobreakdash-\hspace{0pt}convergence topology
does not coincide with the trace norm topology on the set
$\widetilde{D}(\mathcal{A})$, but by noting that the sequences of
eigenvalues of the operators in $\widetilde{D}(\mathcal{A})$ form a
compact subset of the space $l_{1}$ it is easy to see that
$\mu$\nobreakdash-\hspace{0pt}convergence of a sequence
$\{A_{n}\}\subset\widetilde{D}(\mathcal{A})$ to an operator
$A_{0}\in\widetilde{D}(\mathcal{A})$ means trace norm convergence of
the sequence
$\{U_{n}A_{n}U_{n}^{*}\}\subset\widetilde{D}(\mathcal{A})$ to the
operator $A_{0}$ for some set $\{U_{n}\}$ of unitaries.} \textit{If
the quantum entropy is continuous on a compact subset $\mathcal{A}$
of $\,\mathfrak{T}_{+}(\mathcal{H})$ then it is continuous on the
sets $D(\mathcal{A})$ and $\widetilde{D}(\mathcal{A})$ defined in
the first part of assertion E of proposition \ref{UA-property-c}.}
\end{corollary}\vspace{5pt}

This condition and corollary \ref{UA-property-c-c} show that
$$
\left\{H(A_{n}+B_{n})\rightarrow
H(A_{0}+B_{0})\right\}\Leftrightarrow\left\{H(A_{n})\rightarrow
H(A_{0})\right\}\;\wedge\;\left\{H(B_{n})\rightarrow
H(B_{0})\right\},
$$
where $\{A_{n}\}$ and $\{B_{n}\}$ are sequences of positive trace
class operators converging respectively to operators $A_{0}$ and
$B_{0}$. \vspace{5pt}

The above "dominated-type" continuity conditions can be enriched by
the following one.\vspace{5pt}

\begin{corollary}\label{cont-cond++nv}
\textit{If the quantum entropy is continuous on a compact subset
$\mathcal{A}$ of $\,\mathfrak{T}_{+}(\mathcal{H})$, which does not
contain the null operator, then it is continuous on the set
$\widehat{D}(\mathcal{A})$ defined in the second part of assertion E
of proposition \ref{UA-property-c}.}
\end{corollary}\vspace{5pt}

By Corollary \ref{cont-cond++nv} and theorem 13 in \cite{W+} to
prove continuity of the von Neumann entropy on a set
$\mathcal{A}\subset\mathfrak{S}(\mathcal{H})$ it suffice to show its
continuity on the image of this set under the expectation
$\rho\mapsto\sum_{i}P_{i}\rho P_{i}$ for a particular set
$\{P_{i}\}$ of mutually orthogonal projectors such that
$\sum_{i}P_{i}=I_{\mathcal{H}}$.

Corollary \ref{cont-cond++nv} and the infinite-dimensional
generalization of Nielsen's theorem provide the following
observation concerning the notion of entanglement of a state of a
composite quantum system.\vspace{5pt}

\begin{example}\label{entanglement}
Let $\mathcal{H}$ and $\mathcal{K}$ be separable Hilbert spaces. The
entanglement $E(\omega)$ of a pure state $\omega$ in
$\mathfrak{S}(\mathcal{H}\otimes\mathcal{K})$ is defined as the von
Neumann entropy of its reduced states (cf.\cite{B&Co}):
$$
E(\omega)=H(\mathrm{Tr}_{\mathcal{K}}\omega)=H(\mathrm{Tr}_{\mathcal{H}}\omega).
$$
Let $\mathfrak{L}(\mathcal{H},\mathcal{K})$ be the set of all
LOCC\nobreakdash-\hspace{0pt}operations transforming the set
$\mathfrak{S}(\mathcal{H}\otimes\mathcal{K})$ into itself. Corollary
\ref{cont-cond++nv} and lemma 2 in \cite{N}\footnote{In \cite{N} the
majorization order is used, which is converse to the Uhlmann order
$"\prec"$ used in this paper.} imply the following assertion:\\
\textit{If the function $\omega\mapsto E(\omega)$ is continuous on a
compact set
$\,\mathcal{C}\subset\,\mathrm{extr}\mathfrak{S}(\mathcal{H}\otimes\mathcal{K})$
then it is continuous on the set}
$$
\{\Lambda(\omega)\,|\,\omega\in\mathcal{C},\,
\Lambda\in\mathfrak{L}(\mathcal{H},\mathcal{K})\}\cap\mathrm{extr}\mathfrak{S}(\mathcal{H}\otimes\mathcal{K}).
$$
This shows that for arbitrary sequence $\{\omega_{n}\}$ of pure
states in $\mathfrak{S}(\mathcal{H}\otimes\mathcal{K})$ converging
to a state $\omega_{0}$ and arbitrary set $\{\Lambda_{n}\}_{n\geq0}$
of LOCC\nobreakdash-\hspace{0pt}operations such that the sequence
$\{\Lambda_{n}(\omega_{n})\}$ consists of pure states and converges
to the state $\Lambda_{0}(\omega_{0})$ the following implication
holds:
$$
\lim_{n\rightarrow+\infty}E(\omega_{n})=E(\omega_{0})\quad\Rightarrow
\quad\lim_{n\rightarrow+\infty}E(\Lambda_{n}(\omega_{n}))=E(\Lambda_{0}(\omega_{0})).\,\;\square
$$
\end{example}

By corollary \ref{g-mu-comp+} in the Appendix for arbitrary closed
set $\mathcal{A}\subset\mathfrak{T}_{+}(\mathcal{H})$ (not
necessarily compact) and arbitrary natural $m$ the set
$\mathrm{co}_{m}(\mathcal{A})$ defined in assertion D of proposition
\ref{UA-property-c} is closed. Theorem \ref{cont-cond} implies the
following result.\vspace{5pt}
\begin{corollary}\label{co-m}
A) \textit{If the quantum entropy is continuous and bounded on a
closed bounded set $\mathcal{A}\subset\mathfrak{T}_{+}(\mathcal{H})$
then it is continuous on the set $\,\mathrm{co}_{m}(\mathcal{A})$
for arbitrary natural $m$.}\vspace{5pt}

B) \textit{If the quantum entropy is continuous on a compact set
$\,\mathcal{A}\subset\mathfrak{T}_{+}(\mathcal{H})$ then it is
continuous on the set
$\,\mathrm{cl}(\mathrm{co}_{\mathfrak{P}}(\mathcal{A}))$ for
arbitrary subset $\,\mathfrak{P}$ of $\,\mathfrak{P}_{+\infty}$ such
that
$\displaystyle\lim_{m\rightarrow+\infty}\sup_{\{\pi_{i}\}\in\mathfrak{P}}H(\{\pi_{i}\}_{i>m})=0$.}
\end{corollary}\vspace{5pt}

By remark \ref{UA-property-c-r} the set
$\mathrm{cl}(\mathrm{co}_{\mathfrak{P}}(\mathcal{A}))$ in the second
assertion of this corollary can not be replaced by the
$\sigma$\nobreakdash-\hspace{0pt}convex hull
$\sigma\textrm{-}\mathrm{co}(\mathcal{A})$ of the set
$\mathcal{A}$.\vspace{5pt}

\textbf{Proof.} A) Let
$\{A_{n}\}\subset\mathrm{co}_{m}(\mathcal{A})$ be a sequence
converging to an operator $A_{0}\in\mathrm{co}_{m}(\mathcal{A})$.
Suppose
\begin{equation}\label{discont}
\lim_{n\rightarrow+\infty}H(A_{n})>H(A_{0}).
\end{equation}

By the construction of the set $\mathrm{co}_{m}(\mathcal{A})$ for
each $n$ there exists an ensemble $\{\pi^{n}_{i},
A^{n}_{i}\}_{i=1}^{m}$ of operators in $\mathcal{A}$ such that
$A_{n}=\sum_{i=1}^{m}\pi^{n}_{i}A^{n}_{i}$. By proposition
\ref{g-mu-comp} in the Appendix we may consider (by replacing the
sequence $\{A_{n}\}$ by some its subsequence) that there exist
$\lim_{n\rightarrow+\infty}\pi^{n}_{i}=\pi^{0}_{i}$ and
$\lim_{n\rightarrow+\infty}A^{n}_{i}=A^{0}_{i}$ for each
$i=\overline{1,p}$, $p\leq m$,
$\lim_{n\rightarrow+\infty}\sum_{i=1}^{p}\pi^{n}_{i}=1$ and
$A_{0}=\sum_{i=1}^{p}\pi^{0}_{i}A^{0}_{i}$.

For each $n>0$ let
$A'_{n}=\lambda_{n}^{-1}\sum_{i=1}^{p}\pi_{i}^{n}A_{i}^{n}$ is an
operator in $\mathrm{co}_{p}(\mathcal{A})$, where
$\lambda_{n}=\sum_{i=1}^{p}\pi_{i}^{n}$. Since the sequence
$\{\lambda_{n}\}$ tends to $1$ and the set $\mathcal{A}$ is bounded,
the sequence $\{A'_{n}\}$ converges to the operator $A_{0}$. By
theorem \ref{cont-cond}B continuity of the entropy on the compact
set
$$
\mathcal{A}_{*}=\bigcup_{i=1}^{p}\;\{A^{n}_{i}\}_{n\geq
0}\subseteq\mathcal{A}
$$
means the UA\nobreakdash-\hspace{0pt}property of this set. Hence
theorem \ref{cont-cond}C implies continuity of the entropy on the
set $\mathrm{co}_{p}(\mathcal{A}_{*})$ containing the sequence
$\{A'_{n}\}$ and its limit $A_{0}$. Hence
\begin{equation}\label{t-r+}
\lim_{n\rightarrow+\infty}H(A'_{n})=H(A_{0}).
\end{equation}

If $\lambda_{n}=1$ then $A_{n}=A'_{n}$. If $\lambda_{n}<1$ then
$A_{n}=\lambda_{n}A'_{n}+(1-\lambda_{n})A''_{n}$, where
$A''_{n}=(1-\lambda_{n})^{-1}(A_{n}-\lambda_{n}A'_{n})$ is an
operator in $\mathrm{co}_{m}(\mathcal{A})$, and hence by using
inequality (\ref{w-k-ineq}) we obtain
$$
H(A_{n})\leq
\lambda_{n}H(A'_{n})+(1-\lambda_{n})H(A''_{n})+\|A_{n}\|_{1}h_{2}(\lambda_{n}),\quad
\forall n>0.
$$
This shows that (\ref{t-r+}) implies
$\lim_{n\rightarrow+\infty}H(A_{n})=H(A_{0})$ contradicting to
(\ref{discont}) since boundedness of the entropy on the set
$\mathcal{A}$ implies boundedness of the entropy on the set
$\mathrm{co}_{m}(\mathcal{A})$ by inequality (\ref{w-k-ineq}).

B) This directly follows from theorem \ref{cont-cond}. $\square$
\vspace{5pt}

If $\mathcal{A}$ is a union of $\,m<+\infty\,$ closed convex sets
then corollary \ref{g-mu-comp+} in the Appendix implies
$\mathrm{co}_{m}(\mathcal{A})=\overline{\mathrm{co}}(\mathcal{A})$,
so we obtain from corollary \ref{co-m} the following
result.\vspace{5pt}
\begin{corollary}\label{cont-cond+}
\textit{If the quantum entropy is continuous on each set from a
finite collection $\{\mathcal{A}_{i}\}_{i=1}^{m}$ of convex closed
bounded subsets of $\,\mathfrak{T}_{+}(\mathcal{H})$ then it is
continuous on the convex closure
$\overline{\mathrm{co}}\left(\bigcup_{i=1}^{m}\mathcal{A}_{i}\right)$
of this collection.}
\end{corollary}\vspace{5pt}

\begin{remark}\label{cont-cond+r}
The condition of closedness of the \textit{all} sets from the
collection $\{\mathcal{A}_{i}\}_{i=1}^{m}$ in corollary
\ref{cont-cond+} is essential. The simple example showing this can
be constructed as follows. Let $\mathcal{A}_{1}=\{\rho_{0}\}$ and
$\mathcal{A}_{2}=\sigma\textrm{-}\mathrm{co}\left(\{\sigma_{i}\}_{i\in\mathbb{N}}\right)$,
where the state $\rho_{0}$ and the sequence
$\{\sigma_{i}\}_{i\in\mathbb{N}}$ are taken from the example in
remark \ref{ns-cont-cond-r}. As shown in this example the entropy is
continuous on the convex bounded sets $\mathcal{A}_{1}$ and
$\mathcal{A}_{2}$ but it is not continuous on the convex set
$\mathcal{A}_{1}\cup\mathcal{A}_{2}$. $\square$
\end{remark}\vspace{5pt}

Theorem \ref{cont-cond} also implies the following continuity
condition.\vspace{5pt}

\begin{corollary}\label{cont-cond+++}
\textit{Let $\{\mathcal{A}_{i}\}_{i=1}^{n}$ be a finite collection
of subsets of $\,\mathfrak{T}_{+}(\mathcal{H})$ having the
UA\nobreakdash-\hspace{0pt}property (for example, compact subsets on
which the quantum entropy is continuous). Then for arbitrary natural
$m$ the quantum entropy is continuous on the set}
$$
\mathrm{cl}\left(\left\{\sum_{i=1}^{n}\sum_{j=1}^{m}V_{ij}A_{i}V_{ij}^{*}\,|\,A_{i}\in
\mathcal{A}_{i}, V_{ij}\in \mathfrak{B}(\mathcal{H}), \|V_{ij}\|\leq
1\right\}\right).
$$
\end{corollary}

The following observation can be used in study of quantum channels
and in the theory of quantum measurements (see example
\ref{measurement} below).\vspace{5pt}

\begin{corollary}\label{new} \emph{Let $\mathfrak{V}_{=1}$ be the set of all sequences
$\{V_{i}\}_{i=1}^{+\infty}\subset\mathfrak{B}(\mathcal{H})$ such
that $\sum_{i=1}^{+\infty}V^{*}_{i}V_{i}=I_{\mathcal{H}}$ endowed
with the Cartesian product of the strong* operator topology (the
topology of coordinate-wise strong* operator
convergence).\footnote{The strong* operator topology on
$\mathfrak{B}(\mathcal{H})$ is defined by the family of seminorms
\break $A\mapsto\|A|\varphi\rangle\|+\|A^{*}|\varphi\rangle\|$,
$|\varphi\rangle\in\mathcal{H}$ \cite{B&R}. By using more
complicated analysis it is possible to replace this topology here by
the strong operator topology.} Let $\mathcal{A}$ be a subset of
$\,\mathfrak{T}_{+}(\mathcal{H})$ on which the quantum entropy is
continuous.}
\begin{enumerate}[1)]
    \item \emph{The function $\left(\{V_{i}\},A\right)\mapsto
    \sum_{i=1}^{+\infty}H\left(V_{i}AV_{i}^{*}\right)$ is continuous
    on $\,\mathfrak{V}_{=1}\times\mathcal{A}$.}\vspace{5pt}
    \item \emph{If $\,\mathfrak{V}_{0}$ is a subset of
$\,\mathfrak{V}_{=1}$ such that the function
\begin{equation}\label{average-ent}
\left(\{V_{i}\},A\right)\mapsto
H\left(\left\{\mathrm{Tr}V_{i}AV_{i}^{*}\right\}_{i=1}^{+\infty}\right)
\end{equation}
 is continuous on $\,\mathfrak{V}_{0}\times\mathcal{A}$ then the
function $\left(\{V_{i}\},A\right)\mapsto
H\left(\sum_{i=1}^{+\infty}V_{i}AV_{i}^{*}\right) $ is continuous on
$\,\mathfrak{V}_{0}\times\mathcal{A}$.}
\end{enumerate}
\end{corollary}\vspace{5pt}

\textbf{Proof.} We can consider that the sets $\mathcal{A}$ and
$\mathfrak{V}_{0}$ are compact.

1) Corollary \ref{cont-cond+++} shows that the function
$F_{m}(\left(\{V_{i}\},A\right))=H(C_{m}AC_{m})$, where
$C_{m}=\sqrt{\sum_{i=1}^{m}V^{*}_{i}V_{i}}$ and $m\in\mathbb{N}$, is
continuous on $\mathfrak{V}_{=1}\times\mathcal{A}$. Since
$C^{2}_{m}\leq C^{2}_{m+1}$ for all $m$ the sequence $\{F_{m}\}$ is
nondecreasing. By using theorem 1 in \cite{D-A}\footnote{By this
theorem to prove convergence of a sequence
$\{A_{n}\}\subset\mathfrak{T}_{+}(\mathcal{H})$  to an operator
$A_{0}\in\mathfrak{T}_{+}(\mathcal{H})$ it is sufficient to show
that $\lim_{n}\mathrm{Tr}A_{n}=\mathrm{Tr}A_{0}$ and
$\lim_{n}A_{n}=A_{0}$ in the weak operator topology.} and corollary
\ref{cont-cond+++} we conclude that
$\lim_{m\rightarrow+\infty}F_{m}(\left(\{V_{i}\},A\right))=H(A)$.

By the Groenevold-Lindblad-Ozawa inequality (see \cite{O}) we have
$$
\sum_{i>m}H(V_{i}AV^{*}_{i})\leq
H(A)-F_{m}(\left(\{V_{i}\},A\right)).
$$
Hence continuity of the function $A\mapsto H(A)$ and Dini's lemma
show that
$\lim_{m\rightarrow+\infty}\sup_{\{V_{i}\}\in\mathfrak{V}_{c},A\in\mathcal{A}}\sum_{i>m}H(V_{i}AV^{*}_{i})=0$
for arbitrary compact subset $\mathfrak{V}_{c}$ of
$\mathfrak{V}_{=1}$. This and continuity of the function
$\left(\{V_{i}\},A\right)\mapsto
\sum_{i=1}^{m}H\left(V_{i}AV_{i}^{*}\right)$ for each $m$ (provided
by corollary \ref{cont-cond+++}) imply the first assertion of the
corollary.

2) By the above observation the second alternative in condition 1)
in proposition \ref{UA-property-c}F holds for the sets
$\mathfrak{V}_{0}$ and $\mathcal{A}$. Since condition 2) in this
proposition follows from continuity of function (\ref{average-ent})
by Dini's lemma, the set
$\left\{\sum_{i=1}^{+\infty}V_{i}AV_{i}^{*}\,\left|\,\{V_{i}\}\in\mathfrak{V}_{0},
A\in\mathcal{A}\right.\right\}$ has the
UA\nobreakdash-\hspace{0pt}property. By theorem \ref{cont-cond} this
implies the second assertion of the corollary. $\square$
\vspace{5pt}

\begin{example}\label{measurement} Let $\mathfrak{M}_{m}(\mathcal{H})$ be the set
of all quantum measurements with $m\leq+\infty$ outcomes on the
quantum system associated with the Hilbert space $\mathcal{H}$. Each
measurement $\mathcal{M}\in\mathfrak{M}_{m}(\mathcal{H})$ is
described by a set $\{V_{i}\}_{i=1}^{m}$ of operators in
$\mathfrak{B}(\mathcal{H})$ such that
$\sum_{i=1}^{m}V^{*}_{i}V_{i}=I_{\mathcal{H}}$ and its action on
arbitrary a priori state $\rho\in\mathfrak{S}(\mathcal{H})$ results
in the posteriori ensemble $\{\pi_{i}(\mathcal{M},\rho),
\rho_{i}(\mathcal{M},\rho)\}_{i=1}^{m}$, where
$\rho_{i}(\mathcal{M},\rho)=(\mathrm{Tr}V_{i}\rho
V_{i}^{*})^{-1}V_{i}\rho V_{i}^{*}$ is the posteriori
state\footnote{If $\mathrm{Tr}V_{i}\rho V_{i}^{*}=0$ then the
posteriori state $\rho_{i}(\mathcal{M},\rho)$ is not defined.}
corresponding to $i$\nobreakdash-\hspace{0pt}th outcome and
$\pi_{i}(\mathcal{M},\rho)=\mathrm{Tr}V_{i}\rho V_{i}^{*}$ is the
probability of this outcome \cite{H-SSQT}. The mean posteriori state
$\bar{\rho}(\mathcal{M},\rho)=\sum_{i=1}^{m}\pi_{i}(\mathcal{M},\rho)\rho_{i}(\mathcal{M},\rho)=\sum_{i=1}^{m}V_{i}\rho
V_{i}^{*}$ corresponds to the nonselective measurement.

We will consider that a sequence
$\{\mathcal{M}_{n}\}\subset\mathfrak{M}_{m}(\mathcal{H})$ converges
to a measurement $\mathcal{M}_{0}\in\mathfrak{M}_{m}(\mathcal{H})$
if $\,\lim_{n\rightarrow+\infty}V^{n}_{i}=V^{0}_{i}\,$ for all
$\,i=\overline{1,m}\,$ in the strong* operator topology, where
$\{V^{n}_{i}\}_{i=1}^{m}$ is the set of operators describing the
measurement $\mathcal{M}_{n}$.

Let $\mathcal{A}$ be a subset of $\mathfrak{S}(\mathcal{H})$ on
which the von Neumann entropy is continuous. Corollaries
\ref{cont-cond+++} and \ref{new} imply the following assertions:
\begin{itemize}
    \item the von Neumann entropy of posteriory state $H(\rho_{i}(\mathcal{M},\rho))$ is continuous on the subset
    of $\mathfrak{M}_{m}(\mathcal{H})\times\mathcal{A}$, on which $\rho_{i}(\mathcal{M},\rho)$ is defined,  $i=\overline{1,m}\,$;
    \item the mean entropy of posteriory states
    $\sum_{i=1}^{m}\pi_{i}(\mathcal{M},\rho)H(\rho_{i}(\mathcal{M},\rho))$
    is continuous on
    $\,\mathfrak{M}_{m}(\mathcal{H})\times\mathcal{A}$;
    \item if $\mathfrak{M}$ is a subset of $\mathfrak{M}_{m}(\mathcal{H})$ such
    that the Shannon entropy of the outcomes probability
distribution
$H\left(\left\{\pi_{i}(\mathcal{M},\rho)\right\}_{i=1}^{m}\right)$
is continuous on $\,\mathfrak{M}\times\mathcal{A}$ then the von
Neumann entropy of the mean posteriori state
$H\left(\bar{\rho}(\mathcal{M},\rho)\right)$ is continuous on
$\,\mathfrak{M}\times\mathcal{A}$.
\end{itemize}
If $m<+\infty$ then the function
$H\left(\bar{\rho}(\mathcal{M},\rho)\right)$ is continuous on
$\mathfrak{M}_{m}(\mathcal{H})\times\mathcal{A}$. $\square$

\end{example}\vspace{5pt}

\begin{remark}\label{cont-cond-r}
The continuity conditions considered in this subsection are
formulated for subsets of $\mathfrak{T}_{+}(\mathcal{H})$. They can
be  reformulated for subsets of $\mathfrak{S}(\mathcal{H})$ by using
the following obvious observation: \textit{If the quantum entropy is
continuous on a subset $\mathcal{A}$ of
$\,\mathfrak{T}_{+}(\mathcal{H})$ such that
$\,\inf_{A\in\mathcal{A}}\|A\|_{1}>0$ then the von Neumann entropy
is continuous on the subset $\,\left\{
A\|A\|_{1}^{-1}\,|\,A\in\mathcal{A}\right\}$ of
$\,\mathfrak{S}(\mathcal{H})$}.
\end{remark}\vspace{5pt}

\section{Conclusion}

The method of proving continuity of the von Neumann entropy proposed
in this paper is essentially based on the strong stability property
(stated in theorem \ref{s-stability}) and on the $\mu$-compactness
(stated in proposition 2 in \cite{H-Sh-2}) of the set of quantum
states, revealing the special relations between the topology and the
convex structure of this set. Of course, it does not mean that
validity of the continuity conditions obtained by this method
depends on validity of these abstract properties and that these
conditions can not be proved by other methods. For example, the
assertion of corollary \ref{cont-cond+} for sets of quantum states
can be shown by noting that continuity of the entropy on any closed
convex set of states implies compactness of this set (this follows
from lemma 2 in \cite{Sh-9} and corollary 5 in \cite{Sh-4}) and by
applying spectral finite dimensional approximation based on using
inequality (\ref{w-k-ineq}) and Dini's lemma, but the proposed
method provides a more simple and in a sense natural way of doing
this.

The special approximation of concave lower semicontinuous functions
considered in this paper, in particular, the approximation of the
von Neumann entropy used in proving its continuity seems to be
interesting for other applications.

\section{Appendix}

\subsection{One property of the positive cone\\ of trace-class
operators}

The positive cone  $\mathfrak{T}_{+}(\mathcal{H})$  has the
following important property.\vspace{5pt}

\begin{property}\label{g-mu-comp}
\textit{Let $\,\{\{\pi^{n}_{i},A^{n}_{i}\}_{i=1}^{m}\}_{n}$ be a
sequence of ensembles consisting of $m<+\infty$ operators in
$\mathfrak{T}_{+}(\mathcal{H})$ such that the sequence
$\{\sum_{i=1}^{m}\pi^{n}_{i}A^{n}_{i}\}_{n}$ of their averages
converges to an operator $A_{0}$. There exists subsequence
$\{\{\pi^{n_k}_{i},A^{n_k}_{i}\}_{i=1}^{m}\}_{k}$ converging to a
particular ensemble\footnote{We do not assert that $A^{0}_{i}\neq
A^{0}_{j}$ for all $i\neq j$.} $\{\pi^{0}_{i},A^{0}_{i}\}_{i=1}^{m}$
with the average $A_{0}$ in the following sense
$$
\lim_{k\rightarrow+\infty}\pi^{n_k}_{i}=\pi^{0}_{i}\quad\textrm{and}\quad\pi^{0}_{i}>0\;\,\Rightarrow\,
\lim_{k\rightarrow+\infty}A^{n_k}_{i}=A^{0}_{i},\quad
i=\overline{1,m}.
$$}

\end{property}\vspace{5pt}
\textbf{Proof.} Since the set of atomic measures with bounded number
of atoms is weakly closed, it is sufficient to prove that the
sequence $\{\{\pi^{n}_{i},A^{n}_{i}\}_{i=1}^{m}\}_{n}$ considered as
a sequence of measures in
$\mathcal{P}(\mathfrak{T}_{+}(\mathcal{H}))$ is relatively compact.
By Prokhorov theorem (see \cite{Par}) this means that this sequence
is tight, t.i. for arbitrary $\varepsilon>0$ there exists compact
subset $\mathcal{C}_{\varepsilon}$ of
$\mathfrak{T}_{+}(\mathcal{H})$ such that
\begin{equation}\label{d-estimation}
\sup_{n}\left(\sum_{i:A_{i}^{n}\in\mathfrak{T}_{+}(\mathcal{H})\setminus\mathcal{C}_{\varepsilon}}\pi_{i}^{n}\right)<\varepsilon.
\end{equation}

Let $A_{n}=\sum_{i=1}^{m}\pi^{n}_{i}A^{n}_{i}$ for all $n>0$. Since
the set $\{A_{n}\}_{n\geq 0}$ is compact, lemma \ref{g-mu-comp++}
below implies existence of strictly positive
$\mathfrak{H}$\nobreakdash-\hspace{0pt}operator $H$ such that
$\sup_{n}\mathrm{Tr}HA_{n}=c_{0}<+\infty$. Hence for each $n$ and
arbitrary $c>0$ we have
$$
c\sum_{i:\mathrm{Tr}HA_{i}^{n}>c}\pi_{i}^{n}\leq
\sum_{i:\mathrm{Tr}HA_{i}^{n}>c}\pi_{i}^{n}\mathrm{Tr}HA_{i}^{n}\leq
\sum_{i=1}^{m}\pi_{i}^{n}\mathrm{Tr}HA_{i}^{n}=\mathrm{Tr}HA_{n}\leq
c_{0},
$$
which implies
$\sup_{n}\left(\sum_{i:\mathrm{Tr}HA_{i}^{n}>c}\pi_{i}^{n}\right)\leq
c^{-1}c_{0}$. By choosing $c>\varepsilon^{-1}c_{0}$ for given
$\varepsilon>0$ we obtain (\ref{d-estimation}) with the set
$\mathcal{C}_{\varepsilon}=\{A\in\mathfrak{T}_{+}(\mathcal{H})\,|\,\mathrm{Tr}HA\leq
c\}$, which is compact by lemma \ref{g-mu-comp++} below.
$\square$\vspace{5pt}

\begin{corollary}\label{g-mu-comp+}
\textit{For arbitrary closed subset $\mathcal{A}$ of
$\,\mathfrak{T}_{+}(\mathcal{H})$ and arbitrary natural $m$ the set
$\,\mathrm{co}_{m}(\mathcal{A})=\left\{\sum_{i=1}^{m}
\pi_{i}A_{i}\,| \{\pi_{i}\}\in\mathfrak{P}_{m},
\{A_{i}\}\subset\mathcal{A}\right\}$ is closed.}
\end{corollary}\vspace{5pt}

\begin{lemma}\label{g-mu-comp++}
\textit{A closed subset $\mathcal{A}$ of
$\,\mathfrak{T}_{+}(\mathcal{H})$ is compact if and only if there
exists strictly positive
$\mathfrak{H}$\nobreakdash-\hspace{0pt}operator\footnote{see
definition in section 2.} $H$ in $\mathcal{H}$ such that
$\,\sup_{A\in\mathcal{A}}\mathrm{Tr}HA<+\infty$.}
\end{lemma}\vspace{5pt}
\textbf{Proof.} If the set $\mathcal{A}$ is compact then it is
bounded. Hence we may assume that
$\mathcal{A}\subset\mathfrak{T}_{1}(\mathcal{H})$. By using the
compactness criterion for subsets of $\mathfrak{T}_{1}(\mathcal{H})$
(proposition \ref{comp-crit} below) one can construct an increasing
sequence $\{P_{n}\}_{n\geq1}$ of finite rank projectors in
$\mathcal{H}$ strongly converging to the identity operator
$I_{\mathcal{H}}$ such that $\mathrm{Tr}(I_{\mathcal{H}}-P_{n})A\leq
n^{-3}$ for all $A\in\mathcal{A}$. The strictly positive
$\mathfrak{H}$\nobreakdash-\hspace{0pt}operator
$H=P_{1}+\sum_{n=1}^{+\infty}n(P_{n+1}-P_{n})$ has the desired
property.

It is easy to see that existence of \textit{strictly} positive
$\mathfrak{H}$\nobreakdash-\hspace{0pt}operator $H$ such that
$\sup_{A\in\mathcal{A}}\mathrm{Tr}HA<+\infty$ implies boundedness of
the set $\mathcal{A}$. Hence we may assume that
$\mathcal{A}\subset\mathfrak{T}_{1}(\mathcal{H})$. Thus compactness
of the set $\mathcal{A}$ can be proved by using the compactness
criterion for subsets of $\mathfrak{T}_{1}(\mathcal{H})$ mentioned
before and the following inequality
$$
h_{n}\mathrm{Tr}(I_{\mathcal{H}}-P_{n})A\leq \mathrm{Tr}HA,\quad
A\in\mathcal{A},\quad n\in\mathbb{N},
$$
where $h_{n}$ is the $n$-th eigenvalue of the
$\mathfrak{H}$\nobreakdash-\hspace{0pt}operator $H$ (in the
nondecreasing order) and $P_{n}$ is the spectral projector of this
operator corresponding to the eigenvalues $h_{1},...,h_{n}$.
$\square$

The following compactness criterion for subsets of
$\mathfrak{T}_{1}(\mathcal{H})$ can be derived from the compactness
criterion for subsets of $\mathfrak{S}(\mathcal{H})$, presented in
\cite[the Appendix]{H-Sh-2} (by considering the set
$\{A+(1-\mathrm{Tr}A)\rho_{0}\,|\,A\in\mathcal{A}\}\subset\mathfrak{S}(\mathcal{H})$
for a given set $\mathcal{A}\subset\mathfrak{T}_{1}(\mathcal{H})$,
where $\rho_{0}$ is a fixed pure state).\vspace{5pt}

\begin{property}\label{comp-crit}
\textit{A closed subset $\mathcal{A}$ of
$\,\mathfrak{T}_{1}(\mathcal{H})$ is compact if and only if for
arbitrary $\varepsilon>0$ there exists a finite rank projector
$P_{\varepsilon}$ in $\,\mathfrak{B}(\mathcal{H})$ such that
$\mathrm{Tr} (I_{\mathcal{H}}-P_{\varepsilon})A<\varepsilon$ for all
$A\in\mathcal{A}$.}
\end{property}

\subsection{The proofs of the auxiliary results}

\textbf{The proof of Lemma \ref{ca-l-2}.} It is easy to see that
lower semicontinuity and lower boundedness of the function $f$ imply
lower semicontinuity of the functional
\begin{equation}\label{functional}
 \mathcal{P}(\mathcal{A})\ni\mu\mapsto F(\mu)=\int_{\mathcal{A}}f(\sigma)\mu(d\sigma).
\end{equation}

A) Convexity of the function $\check{f}_{\mathcal{A}}$ follows from
its definition. By lower semicontinuity of the functional
(\ref{functional}) and compactness of the set
$\mathcal{P}_{\{\rho\}}(\mathcal{A})$ (provided by $\mu$-compactness
of the set $\mathfrak{S}(\mathcal{H})$) the infimum  in the
definition of the value $\check{f}_{\mathcal{A}}(\rho)$ for each
$\rho$ in $\overline{\mathrm{co}}(\mathcal{A})$ is achieved at a
particular measure in $\mathcal{P}_{\{\rho\}}(\mathcal{A})$.

Suppose the function $\check{f}_{\mathcal{A}}$ is not lower
semicontinuous. Then there exists a sequence
$\{\rho_{n}\}\subset\overline{\mathrm{co}}(\mathcal{A})$ converging
to a state $\rho_{0}\in\overline{\mathrm{co}}(\mathcal{A})$ such
that
\begin{equation}\label{l-s-b}
\lim\limits_{n\rightarrow+\infty}\check{f}_{\mathcal{A}}(\rho_{n})<\check{f}_{\mathcal{A}}(\rho_{0}).
\end{equation}
As proved before for each $n=1,2,...$ there exists measure $\mu_{n}$
in  $\mathcal{P}_{\{\rho_{n}\}}(\mathcal{A})$ such that
$\check{f}_{\mathcal{A}}(\rho_{n})=F(\mu_{n})$. $\mu$-compactness of
the set $\mathfrak{S}(\mathcal{H})$ implies existence of a
subsequence $\{\mu_{n_{k}}\}$ converging to a particular measure
$\mu_{0}$. By continuity of the map $\mu\mapsto\mathbf{b}(\mu)$ the
measure $\mu_{0}$ belongs to the set
$\mathcal{P}_{\{\rho_{0}\}}(\mathcal{A})$. Lower semicontinuity of
functional (\ref{functional}) implies
$$
\check{f}_{\mathcal{A}}(\rho_{0})\leq
F(\mu_{0})\leq\liminf_{k\rightarrow+\infty}F(\mu_{n_{k}})=\lim_{k\rightarrow+\infty}\check{f}_{\mathcal{A}}(\rho_{n_{k}}),
$$
contradicting to (\ref{l-s-b}).\vspace{5pt}

B) Concavity of the function $\hat{f}_{\mathcal{A}}$ follows from
its definition.

Suppose the function $\hat{f}_{\mathcal{A}}$ is not lower
semicontinuous. Then there exists a sequence
$\{\rho_{n}\}\subset\overline{\mathrm{co}}(\mathcal{A})$ converging
to a state $\rho_{0}\in\overline{\mathrm{co}}(\mathcal{A})$ such
that
\begin{equation}\label{l-s-b+}
\lim\limits_{n\rightarrow+\infty}\hat{f}_{\mathcal{A}}(\rho_{n})<\hat{f}_{\mathcal{A}}(\rho_{0}).
\end{equation}
For $\varepsilon>0$ let $\mu_{0}^{\varepsilon}$ be such measure in
$\mathcal{P}_{\{\rho_{0}\}}(\mathcal{A})$ that
$\hat{f}_{\mathcal{A}}(\rho_{0})<F(\mu_{0}^{\varepsilon})+\varepsilon$.
By openness of the map
$\mathcal{P}(\mathcal{A})\ni\mu\mapsto\mathbf{b}(\mu)$ there exists
a subsequence $\{\rho_{n_{k}}\}$ and a sequence
$\{\mu_{k}\}\subset\mathcal{P}(\mathcal{A})$ converging to the
measure $\mu_{0}^{\varepsilon}$ such that
$\mathbf{b}(\mu_{k})=\rho_{n_{k}}$ for each $k$. Lower
semicontinuity of functional (\ref{functional}) implies
$$
\hat{f}_{\mathcal{A}}(\rho_{0})\leq
F(\mu_{0}^{\varepsilon})+\varepsilon\leq\liminf\limits_{k\rightarrow+\infty}F(\mu_{k})+\varepsilon\leq\lim\limits_{k\rightarrow+\infty}\hat{f}_{\mathcal{A}}(\rho_{n_{k}})+\varepsilon,
$$
contradicting to (\ref{l-s-b+}) (since $\varepsilon$ is arbitrary).
$\square$\vspace{5pt}

\textbf{The proof of the assertion in remark \ref{ns-cont-cond-r}.}
For arbitrary state $\rho$ in
$\mathcal{A}=\sigma\textrm{-}\mathrm{co}(\{\sigma_{i}\})$ there
exists a probability distribution
$\{\pi_{i}\}\in\mathfrak{P}_{+\infty}$ such that
$\rho=\sum_{i=1}^{+\infty}\pi_{i}\sigma_{i}$. This distribution is
unique since $P_{i}\rho=\pi_{i}\lambda_{i}\rho_{i}$ for each $i$,
where $P_{i}$ is the projector on the subspace
$\mathrm{supp}\rho_{i}$.

The one-to-one correspondence
$\mathfrak{P}_{+\infty}\ni\{\pi_{i}\}\leftrightarrow\sum_{i}\pi_{i}\sigma_{i}\in\mathcal{A}$
is continuous in the both directions (t.i. it is a homeomorphism).
Indeed, continuity of the map $"\rightarrow"$ is obvious while
continuity of the map $"\leftarrow"$ can be proved by using the
above set $\{P_{i}\}$ of projectors and by noting that pointwise
convergence of a sequence of probability distributions to a
\textit{probability distribution} implies its convergence in the
norm of total variation.

Thus to prove continuity of the von Neumann entropy on the set
$\mathcal{A}$ it is sufficient to show continuity of the function
$\mathfrak{P}_{+\infty}\ni\{\pi_{i}\}\mapsto
H\left(\sum_{i}\pi_{i}\sigma_{i}\right)$.

By the construction of the sequence $\{\sigma_{i}\}$ we have
$$
\begin{array}{c}
\displaystyle
H\left(\sum_{i=1}^{+\infty}\pi_{i}\sigma_{i}\right)=H\left(\left(\sum_{i=1}^{+\infty}\pi_{i}(1-\lambda_{i})\right)\rho_{0}\oplus
\bigoplus_{i=1}^{+\infty}\pi_{i}\lambda_{i}\rho_{i}\right)\\\\
\displaystyle=\sum_{i=1}^{+\infty}\pi_{i}(1-\lambda_{i})H(\rho_{0})+\sum_{i=1}^{+\infty}\pi_{i}\lambda_{i}H(\rho_{i})
+\eta\left(\sum_{i=1}^{+\infty}\pi_{i}(1-\lambda_{i})\right)
+\sum_{i=1}^{+\infty}\eta(\pi_{i}\lambda_{i})\\\\
\displaystyle
=1+\eta\left(\sum_{i=1}^{+\infty}\pi_{i}(1-\lambda_{i})\right)
+\sum_{i=1}^{+\infty}\pi_{i}\lambda_{i}(-\log\pi_{i})+\sum_{i=1}^{+\infty}\pi_{i}\lambda_{i}(-\log\lambda_{i}).
\end{array}
$$

By using properties of the function $x\mapsto\eta(x)$ and lemma
\ref{simple} below it is easy to show continuity of the all terms in
the right side of the above expression as functions of
$\{\pi_{i}\}$.

Discontinuity of the  von Neumann entropy on the set
$\mathrm{cl}\,(\mathcal{A})=\mathcal{A}\cup\{\rho_{0}\}$ follows
from the inequality $H(\sigma_{i})\geq\lambda_{i}H(\rho_{i})=1,\,
i>0$, since $H(\rho_{0})=0$.\vspace{5pt}

\begin{lemma}\label{simple}
\textit{Let $\{h_{j}\}_{j=1}^{+\infty}$ be a nondecreasing sequence
of positive numbers such that
$\,\mathrm{g}\left(\{h_{j}\}\right)=\inf\left\{\lambda>0\,|\,\sum_{j=1}^{+\infty}e^{-\lambda
h_{j}}<+\infty\right\}<+\infty$. Then
$$
\lim_{m\rightarrow+\infty}\sup_{\{x_{j}\}\in\mathcal{B}_{1}}\sum_{j\geq
m}\eta(x_{j})h^{-1}_{j}=\mathrm{g}\left(\{h_{j}\}\right),
$$
where $\mathcal{B}_{1}$ is the positive part of the unit ball of the
Banach space $\,l_{1}$.}
\end{lemma}\vspace{5pt}

\textbf{Proof.} We will prove first that
\begin{equation}\label{double-ineq}
\lambda_{*}\leq\sup_{\{x_{j}\}\in\mathcal{B}_{1}}\sum_{j=1}^{+\infty}\eta(x_{j})h^{-1}_{j}\leq\lambda_{*}+h^{-1}_{1},
\end{equation}
where $\lambda_{*}$ is either the unique solution of the equation
$\sum_{j=1}^{+\infty}e^{-\lambda h_{j}}=e$ if it exists or
$\mathrm{g}\left(\{h_{j}\}\right)$ otherwise.\footnote{The equation
$\,\sum_{j=1}^{+\infty}e^{-\lambda h_{j}}=e$ has no solution if and
only if $\,\sum_{j=1}^{+\infty}e^{-\mathrm{g}\left(\{h_{j}\}\right)
h_{j}}<e$.}

By using the Lagrange method it is easy to show that the function
$\{x_{j}\}_{j=1}^{n}\mapsto\sum_{j=1}^{n}\eta(x_{j})h^{-1}_{j}$
attains its maximum on the positive part $\mathcal{B}^{n}_{1}$ of
the unit ball of $\mathbb{R}^{n}$ at the vector
$\{e^{-\lambda_{n}h_{j}-1}\}_{j=1}^{n}$, where $\lambda_{n}$ is the
unique solution of the equation $\sum_{j=1}^{n}e^{-\lambda h_{j}}=e$
and hence
$$
\lambda_{n}\leq\sup_{\{x_{j}\}\in\mathcal{B}^{n}_{1}}\sum_{j=1}^{n}\eta(x_{j})h^{-1}_{j}=\lambda_{n}+\sum_{j=1}^{n}e^{-\lambda_{n}
h_{j}-1}h^{-1}_{j}\leq \lambda_{n}+h^{-1}_{1}.
$$
It is easy to see that the increasing sequence $\{\lambda_{n}\}$
converges to $\lambda_{*}$, so by noting that
$\{x_{j}\}\mapsto\sum_{j=1}^{+\infty}\eta(x_{j})h^{-1}_{j}$ is a
lower semicontinuous function and by passing to the limit in the
above expression we obtain (\ref{double-ineq}).

The assertion of the lemma can be derived from (\ref{double-ineq})
applied to the sequence $\,\{h_{j+m}\}_{j=1}^{+\infty}\;$ since if
the solution of the equation $\sum_{j=1}^{+\infty}e^{-\lambda
h_{j+m}}=e\,$ exists for all $m$ it tends to
$\,\mathrm{g}\left(\{h_{j}\}\right)\,$ as $\,m\rightarrow+\infty$.
$\square$\vspace{15pt}

I am grateful to A.S.Holevo for the help and useful discussion. I am
also grateful to I.Nechita for the information about the
generalization of Nielsen's theorem and to the referees for the
useful remarks. This work is partially supported by the program
"Mathematical control theory" of Russian Academy of Sciences, by the
federal target program "Scientific and pedagogical staff of
innovative Russia", project NK-13P(2), by the analytical
departmental target program "Development of scientific potential of
the higher school 2009-2010", project 2.1.1/500 and by RFBR grant
09-01-00424-a. I am grateful to the organizers of the workshop
\emph{Thematic Program on Mathematics in Quantum Information} at the
Fields Institute, where some of this work was done.


\begin{thebibliography}{99}
\bibitem{Max} Maximality of a measure $\mu$ in
$\mathcal{P}(\mathfrak{S}_{1}(\mathcal{H}))$ with respect to the
Choquet ordering follows from coincidence of this ordering with the
dilation ordering \cite{Edgar}, but it can be easily proved in this
special case as follows. Suppose $\nu\succ\mu$. Since for arbitrary
concave continuous bounded function $f$ on the
$\mathfrak{S}(\mathcal{H})$ stability of the set
$\mathfrak{S}(\mathcal{H})$ implies continuity of its convex hull
$\mathrm{co}f$ (see \cite[theorem 1]{P-Sh}) we have
$$
\mu(f)\geq\nu(f)\;\;\textrm{and}\;\;
\mu(\mathrm{co}f)\leq\nu(\mathrm{co}f),\;\; \textrm{where}\;\;
\mu(f)= \int_{\mathfrak{S}(\mathcal{H})}f(\sigma)\mu(d\sigma).
$$

By noting that $f\geq\mathrm{co}f$ and that these functions coincide
on the support of the measure $\mu$ we conclude that $\mu(f)=\nu(f)$
and hence $\mu=\nu$.
\end{thebibliography}
\end{document}